\documentclass[aps,pre,reprint,superscriptaddress]{revtex4-1}
\pdfoutput=1
\usepackage{graphicx}
\usepackage[T1]{fontenc}
\newcommand{\choosefont}[1]{\fontfamily{#1}\selectfont}
\usepackage{amsmath}
\usepackage{float}
\usepackage{slashbox}
\usepackage{textcomp}
\usepackage{hyperref}

\usepackage{multirow}

\begin{document}

\title{Probing Hybridization parameters from microarray experiments:
nearest neighbor model and beyond}

\author{%
W.W. Hadiwikarta\,$^{1,2}$,
J.-C. Walter\,$^{2}$,
J. Hooyberghs\,$^{1,3}$,
E. Carlon$^{2}$
}

\email[To whom correspondence should be addressed:\\Tel: +32 16 32 72 39\\ Fax: +32 16 32 79 86\\Email: ] {enrico.carlon@fys.kuleuven.be}

\address{
Flemish Institute for Technological Research, VITO, Boeretang 200, B-2400 Mol, Belgium
,
$^{2}$Institute for Theoretical Physics, KULeuven, Celestijnenlaan 200D, B-3001 Leuven, Belgium,
and
$^{3}$Department WNI, Hasselt University, Campus Diepenbeek, Agoralaan - Building D, B-3590, Diepenbeek, Belgium}

\begin{abstract}
In this article it is shown how optimized and dedicated microarray
experiments can be used to study the thermodynamics of DNA hybridization
for a large number of different conformations in a highly parallel
fashion. In particular, free energy penalties for mismatches are obtained
in two independent ways and are shown to be correlated with values from
melting experiments in solution reported in the literature. The additivity principle,
which is at the basis of the nearest-neighbor model, and according to 
which the penalty for two isolated mismatches is equal to the sum of 
the independent penalties, is thoroughly tested. Additivity is shown to 
break down for a mismatch distance below $5$ nt.
The behavior of mismatches in the vicinity of the helix edges,
and the behavior of tandem mismatches are also investigated. Finally, some
thermodynamic outlying sequences are observed and highlighted.
These sequences contain combinations of GA mismatches. The analysis of the microarray
data reported in this article provides new insights on the DNA hybridization
parameters and can help to increase the accuracy of hybridization-based
technologies.
\newline
\newline
\end{abstract}

\maketitle

\section{Introduction}

Hybridization of single-stranded nucleic acids to form a duplex
is a reversible chemical reaction, which is at the basis of many
processess and techniques currently used in biotechnology, as
for instance PCR~\cite{albe02}.  Due to its central importance,
hybridization has been intensively studied in experiments (focusing
on the thermodynamics~\cite{bres86,SantaLucia1998} or kinetics of the process)
and also in computer simulations~\cite{Sambriski2009}.

The thermodynamics of DNA hybridization is usually described by
the nearest-neighbor (NN) model~\cite{bloo00}. This model assumes
that the free energy of a duplex can be expressed as a sum of
dinucleotide stability parameters; it is therefore based on the principle of
additivity. From the NN parameters one can, for instance, estimate melting
temperatures, compute melting curves and predict secondary structures
in which RNA molecules fold~\cite{Mathews2004,Andronescu2010}. In the folding problem,
many different local conformations arise as single nucleotide mismatches,
bulges, stem-loop structures, etc. Describing these conformations in
the framework of the NN model is very challenging and requires a large
number of parameters~\cite{Mathews2004}. However, only a limited number of
them have been measured directly in experiments~\cite{SantaLucia2004}.  In addition, one may
also wonder whether additivity holds in such cases. To investigate a
large number of different conformations, it would be very advantageous
to have access to high-throughput measurements, provided that they are
sufficiently accurate.

In this article, we quantitatively determine free energy penalties for
mismatches using microarray data obtained from a set of optimized
and dedicated experiments. In DNA microarrays, several thousand of
different sequences can be spotted at a surface, hence a large number
of hybridization reactions takes place simultaneously.
We use two different approaches: the first one is based on a linear
regression of a large set of experimental data points ($\approx 1000$)
to fit $58$ NN dinucleotide parameters.  The second method relies on the
computation of the logarithm of the ratios of fluorescent intensities
measured from different spots of the arrays. We show that both methods
provide highly correlated set of NN parameters.  In addition, the second
approach allows to probe the limitations of the NN model. It is found
that when two mismatches are closer than $5$ nt additivity
breaks down and the free energy of the duplex is not equal to the sum
of the two separate contributions of isolated mismatches. We also quantify
the influence of mismatches close to the edge of the double helix and
show that the free energy penalty is much weaker in those cases.
Overall, this work provides new insights on DNA hybridization thermodynamics
and can help to increase the accuracy of hybridization-based technologies.

\begin{table*}[!th]
\begin {center}
\begin{tabular}{|cc|}
\hline
$t_1$:   &\choosefont{pcr}5'-CTGGTCTTAGATGC\textbf{AGC}GACTGTTT-poly(A)-3'-Cy3\\
\hline
$t_2$:   &\choosefont{pcr}5'-CTGCACAATTCCGG\textbf{AGC}TATGAATT-poly(A)-3'-Cy3\\
\hline
$t_3$:   &\choosefont{pcr}5'-AATAATGCTCATTAGGCACCGGGAA-poly(A)-3'-Cy3\\
\hline
\end{tabular}
\end{center}
\caption{Target sequences used in the experiments. At the 3' side of
each sequence a 20-mer poly(A) is attached, terminating with a Cy3
fluorophore. The targets were selected from Optimal Design criteria~\cite{atki92}
(Supplementary Data). Each target is hybridized separately on specific microarrays
containing mismatched probes with up to two mismatches with respect to the target.
Note that $t_1$ and $t_2$ share a common triplet of nucleotides AGC at the same sequence
position (in bold characters). The mismatches centered around this
triplet will be discussed in some details in the 'Results' section.
}
\label{TAB01}
\end{table*}

\section{MATERIALS AND METHODS}
\label{sec:setup}

The experiments were performed on custom Agilent arrays, following
a standard protocol, which is  discussed in ~\cite{Hooyberghs2009}. In
each experiment, a single target sequence in solution was hybridized
at concentrations ranging typically from $\sim 10$~picoM to $\sim
2$~nanoM. In total, three different sets of experiments were performed
using the target sequences shown in Table~\ref{TAB01}. These sequences
were selected from $25$-mers human DNAs using Optimal Design methods~\cite{atki92}.
The theory of Optimal Design provides some criteria of selecting an
optimal set of measurements, which minimize the uncertainties in the
parameters of a statistical model (see Supplementary Data).

From the targets of Table~\ref{TAB01}, three different microarrays
were designed and used for hybridization to either $t_1$, $t_2$ or
$t_3$. Each microarray contains probes with either zero, one or two
mismatches with respect to the given target, covering all possible
mismatch combinations. In a stretch of $N$ nucleotides there can be
$3N$ single mismatch probes and $9N(N-1)/2$ double mismatch probes.
For $N=25$ this gives in total $2776$ different sequences, which
were spotted in the microarray. The sequences were replicated $15$
times to fill up completely a $44$K custom Agilent array. Another
design was also used mismatches have a minimal distance of
4 nt from the border and a minimal relative distance of 5 nt.
In this case the total number of sequences is $646$. These sequences were replicated
$23$ times to fill a 15K custom arrays. We considered hybridizing
sequences of 25 nucleotides. This is because in previous
studies~\cite{Hooyberghs2010} these sequences were found to attain
thermodynamic equilibrium after $\sim 3$ h of hybridization (in
the experiments the hybridization time is of $17$ h, hence
thermodynamic equilibrium is guaranteed). A hybridization experiment
provides a large number of fluorescence intensities: the highest
intensity is from spots containing perfect match sequence, whereas the
intensity decreases with the number and type of mismatches.  The
reduction of the intensity provides an estimate of the hybridization
free energy. We use two different methods to obtain the NN
parameters, as discussed in the next sections.

\section{RESULTS}

\subsection{Nearest-neighbor parameters from linear regression}
\label{sec:lin_fit}

Equilibrium thermodynamics predicts that the measured fluorescence
intensity from a spot $i$ equals to:
\begin{eqnarray}
I_i = I_0 + A c e^{-\Delta G_i/RT}
\label{IivsDG}
\end{eqnarray}
where $\Delta G_i$ is the hybridization free energy between the target
sequence and a probe sequence in $i$, $A$
is a parameter, which sets the intensity scale, $c$
the target concentration, $R$ the gas constant and $T$ the
temperature (experiments are performed at $T=65^\circ$C $= 338 K$,
which is the value of the temperature used in the rest of the analysis).
Although the data analyzed are background-subtracted
from the Agilent scanner, there remains always some small aspecific
signals, which we denote by $I_0$ in Equation~(\ref{IivsDG}).  In the
experiments $I_i$ is obtained from the average over typically approximately
$15$ replicated spots. One should note that Equation~(\ref{IivsDG})
is valid at sufficiently low target concentrations, i.e. when only a
limited fraction of probes is hybridized in a spot, hence far from
chemical saturation. On the other hand, at very low concentrations,
the specific signal, i.e. the second term in Equation~(\ref{IivsDG}), can
become comparable to $I_0$.  Therefore, for the analysis of the data
we restricted ourselves to intermediate concentrations and
intensities for which we explicitly verified that the intensities
scale linearly with concentrations, as predicted by
Equation~(\ref{IivsDG}) (more details can be found in the Supplementary
Data). In the intensity scale of the experiments $I_0 \approx 1$,
whereas the values used in the analysis are $I_i \gtrsim 10$.
In practice, the large majority of the intensities
in experiments with target concentration $c=100$~pM or higher are above
this threshold value.

In the following, we will consider the logarithm of the intensities
measured with respect to the perfect match (PM) intensity. Using
Equation~(\ref{IivsDG}), for $I_i \gg I_0$  we get:
\begin{eqnarray}
y_i \equiv \ln I_i - \ln I_{PM} = -\frac{\Delta G_i - \Delta
G_{PM}}{RT} \label{yi_linear}
\end{eqnarray}
which defines the free energy penalty of probe $i$ with respect to
the perfectly matching probe. This penalty can be expressed as a sum
of NN dinucleotide parameters. Consider, for instance, the example of
a probe $i$ with a single mismatch of type A with respect to the
target nucleotide G and with neighboring nucleotides G and T. We
have:
\begin{eqnarray}
\Delta G_i \left( {\ldots G\underline{A}T \ldots}\atop{\ldots
C{\underline G}A \ldots} \right) &-& \Delta G_{PM} \left( {\ldots
GCT \ldots}\atop{\ldots CGA \ldots} \right) =
\nonumber\\
\Delta G \left( {G\underline{A}}\atop{C{\underline G}} \right) &+&
\Delta G \left( {\underline{A}T}\atop{{\underline G}A} \right)
\nonumber\\
- \Delta G \left( {GC}\atop{CG} \right) &-& \Delta G \left(
{CT}\atop{GA} \right)
\nonumber\\
\equiv \Delta \Delta G \left( {G\underline{A}T}\atop{C{\underline
G}A} \right) && \label{yi_eq2}
\end{eqnarray}
We use the following notation: the target sequence is the bottom strand and the probe
sequence, which is oriented from 5'$-$3', is the top strand. This example corresponds to
target $t_{1}$ or $t_{2}$  at position 10, counting from 3' end (the
triplet of nucleotides are indicated in bold in Table~\ref{TAB01}). In
Equation~(\ref{yi_eq2}) $\Delta \Delta G$ is defined as the free energy penalty
of an isolated mismatch in a DNA duplex.  This penalty is expected to
be a local effect. In the NN model this locality is inherent:
the dots in Equation~(\ref{yi_eq2}) indicate identical nucleotides in the two
sequences, their contribution cancels out and leaves per isolated mismatch
only four dinucleotide parameters around the mismatch position. There
are in total only 58 such dinucleotide parameters: $10$ perfect match
parameters and $48$ single mismatch parameters (taking into account
symmetries). The dinucleotide parameters are not directly experimentally
accessible and are not unique~\cite{Gray1997}, e.g.  they can be shifted by
some constant value such that the physically accessible $\Delta \Delta G$
remains unchanged (see Supplementary Data).

\begin{figure}[t!]
\includegraphics[width=1\columnwidth]{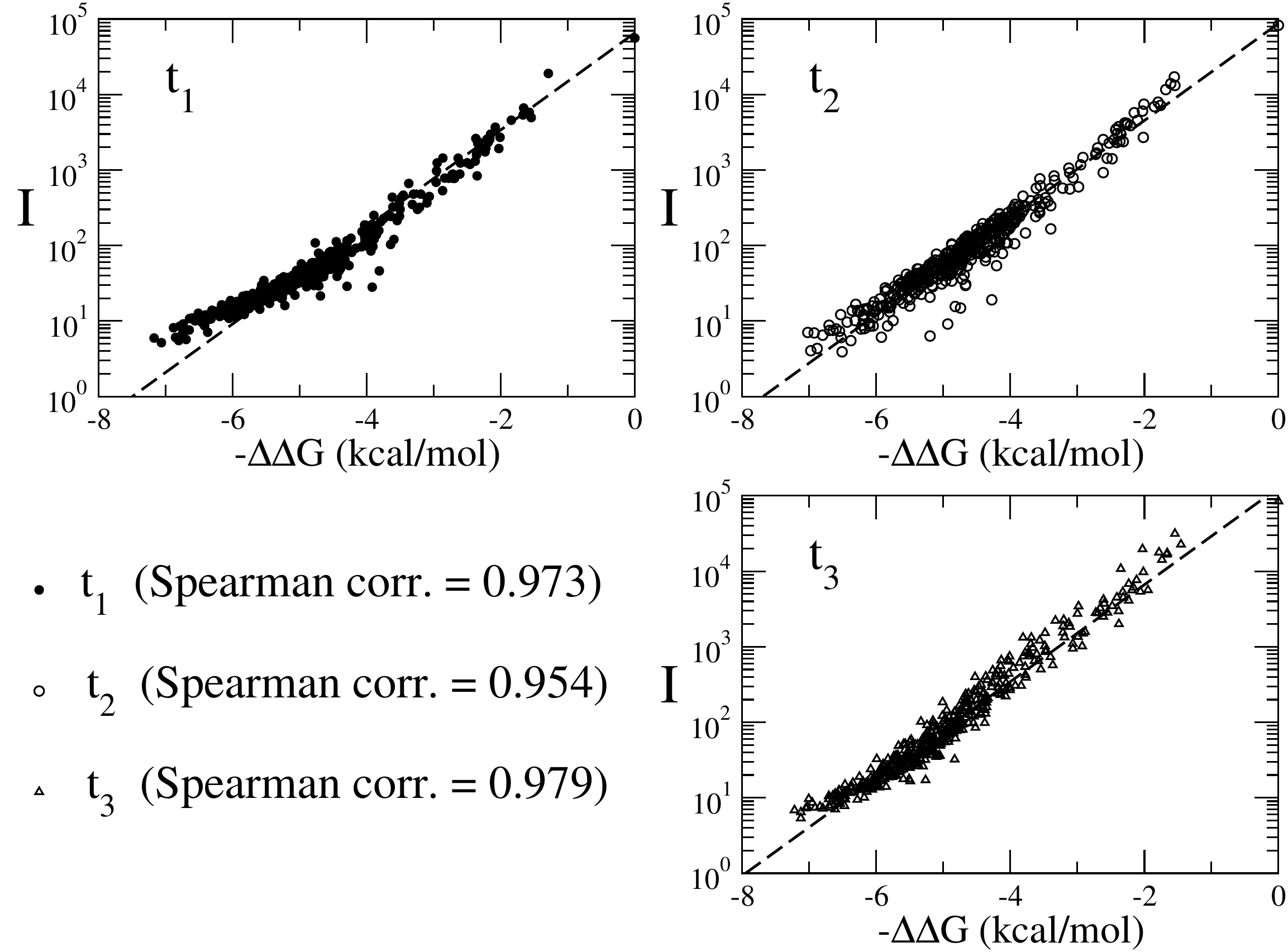}
\caption{Plot of the Intensities for concentrations $c=100$ pM from
the experiments using hybridization of targets $t_1$, $t_2$ or $t_3$,
as a function of the $\Delta \Delta G$ parameters obtained from
least-squared minimization. The data agree well with hybridization
isotherm given in Equation~(\ref{yi_linear}), shown as a straight line
in the linear-log scale.}
\label{FIG_FIT_COLLAPSE}
\end{figure}

Equations~(\ref{yi_linear}) and~(\ref{yi_eq2}) define a linear
problem: each measured $y_i$ can be expressed by a linear
combination of dinucleotide parameters. In order to extract the
parameters from the data we combined the results of the three
experiments and performed a least square minimization of
Equation~(\ref{yi_linear}). Mismatches closer than five sites from the
helix edges were excluded from the analysis, as well as pairs of
mismatches with a distance smaller than 5 nt.

The $58$ adjustable parameters were fitted on a set of about a
thousand of experimental data points above the intensity 
threshold. The fitted parameters then applied to produce the plot 
as shown in Figure~\ref{FIG_FIT_COLLAPSE} for all available intensities
of the experiments in which either sequence $t_1$, $t_2$ or $t_3$
was hybridized on its corresponding microarray at a concentration
of $c=100$~pM. The data are plotted as a function of the unique 
$\Delta \Delta G$ for triplets defined as in Equation~(\ref{yi_eq2}).
We note that there is very good agreement between the data
and the thermodynamic model of Equation~(\ref{IivsDG}).
The experiments follow the equilibrium isotherm (a straight line
with a slope equal to $1/RT$) for a range of intensities of more
than four orders of magnitude.  A previous
study~\cite{Hooyberghs2009} in which hybridizing strands were
30-mers did not provide a single straight line in a $\ln I$ versus
$\Delta \Delta G$ plot. Deviations due to lack of thermodynamic
equilibrium were observed in the high-intensity ranges, as discussed
in~\cite{Hooyberghs2010,Walter2011}.

\begin{figure}[t!]
\includegraphics[width=1\columnwidth]{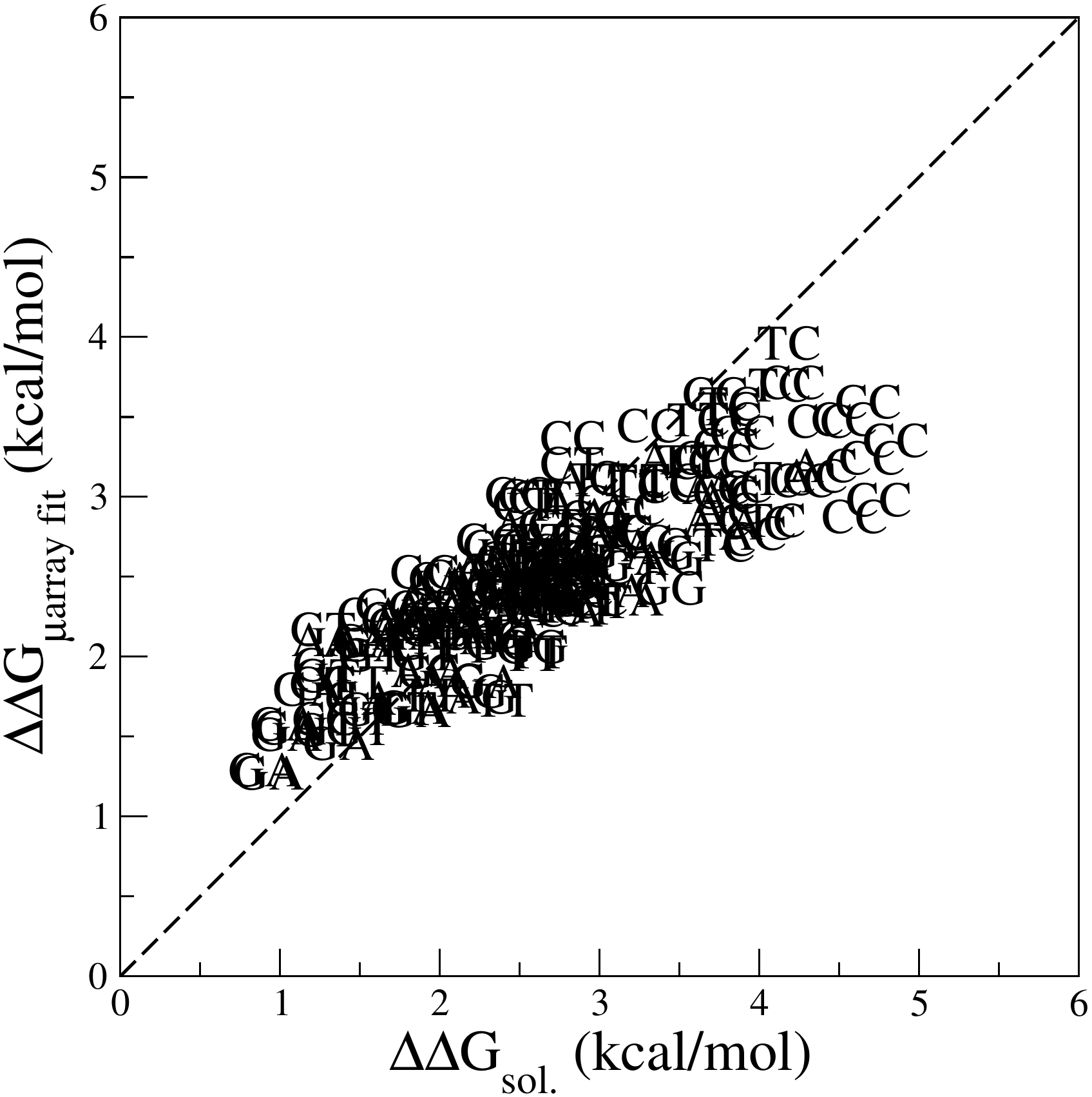}
\caption{Plot of free energy penalties $\Delta \Delta G$ for
triplets obtained from the microarray fit versus those from
hybridization in solution~\cite{SantaLucia2004}. The central
mismatching nucleotides of the triplet (underlined in Equation~(\ref{yi_eq2})) are indicated in the plot.}
\label{FIG_ARRAY_VS_SOLUTION}
\end{figure}

Further it is important to note that we do not only find internally
consistent results, but that our microarray-derived free energy
parameters also correlate to a fair degree with those reported in
literature for hybridization in solution~\cite{SantaLucia2004}.
Figure~\ref{FIG_ARRAY_VS_SOLUTION} shows a correlation plot of the
free energy penalties (i.e. the $\Delta\Delta G$ defined as in the
example of Equation~(\ref{yi_eq2})) obtained from the microarray data analysis and
those from SantaLucia et al. from~\cite{SantaLucia2004}.  The
Spearman correlation coefficient is equal to $0.855$. This clearly
shows that free energy parameters for DNA features measured by the
presented microarray approach also apply for thermodynamic
properties in solution. This opens the highly parallelled microarray
toolbox for the study of thermodynamics of DNA structures. An example is
discussed in the next section.

\subsection{Nearest-neighbor parameters from ratios of intensities: probing additivity}
\label{sec:additivity}

The crucial assumption of the NN model is additivity of local free energy
contributions. We probe here the limits of additivity of free
energy penalties as a function of the distance between two mismatches.
We will access the free energy parameters by comparing ratios of
intensities measured from different spots in the microarray.

\begin{figure}[!t]
\includegraphics[width=1\columnwidth]{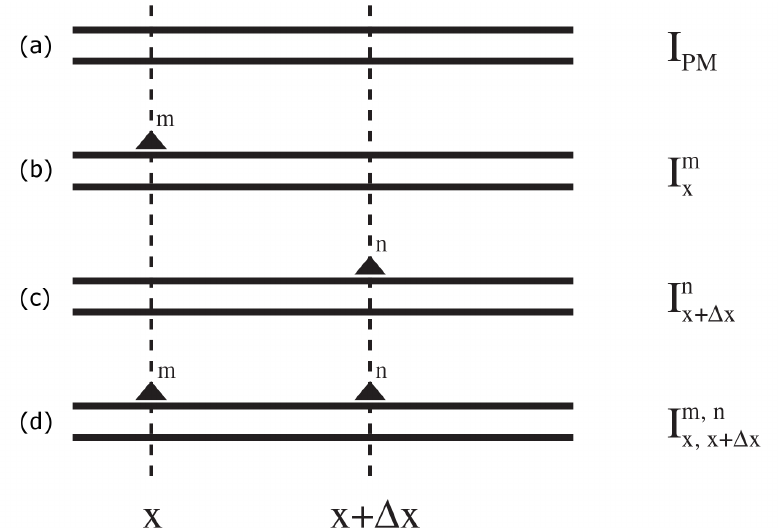}\\
\caption{Schematic representation of hybridizing strands in the microarray
experiment. From the appropriate ratios of intensities measured from
these spots, the free energy parameters can be determined and the additivity
principle can be tested. As in the rest of the article the lower strand
is the fixed target sequence. The upper strand is the probe sequence.
The filled triangles denote mismatching nucleotides. In the four
examples from the top we show: (a) hybridization with a PM probe, (b,c)
hybridization with a single mismatch probe where the
mismatching nucleotides are $m$ and $n$ at positions $x$ and $x+\Delta
x$ respectively, (d) hybridization with a probe carrying two mismatches.
We use the notations $I_x^m$, $I_{x+\Delta x}^n$
and $I_{x,x+\Delta x}^{m,n}$ to denote the corresponding intensities
measured in the experiment.}
\label{FIG_MM_sketch}
\end{figure}

\begin{figure}[t]
  \includegraphics[width=1\columnwidth]{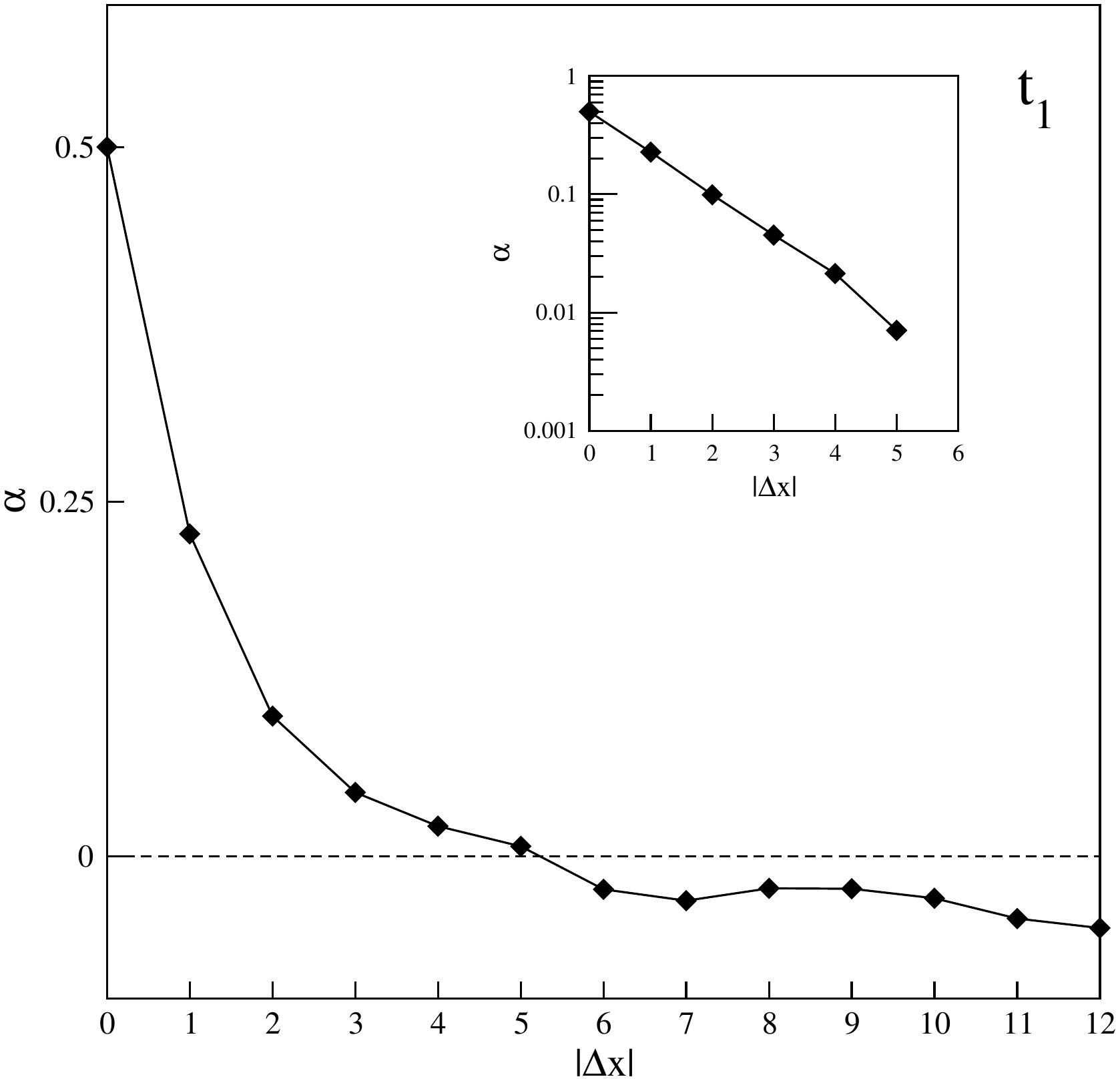}\\
  \caption{Parameter $\alpha$, the relative deviation from additivity, from the experiment of target $t_{1}$,
  averaged over $x, m$ and $n$ as a function of the distance $|\Delta x|$ between two mismatches.
  The inset shows the plot with $\alpha$ in log scale.}\label{FIG_interaction_alpha}
\end{figure}

Hereto, we combine microarray spots that contain probes with zero,
one or two mismatches with respect to the target and we denote the
location of the mismatch by $x$ or $x+ \Delta x$ as illustrated in
Figure~\ref{FIG_MM_sketch}. The associated free energy penalties can
then be derived from the intensity measurements as follows
\begin{eqnarray}
  \Delta\Delta G_{x}^{m} &=& -RT \ln\left(\frac{I^{m}_{x}}{I_{PM}}\right) \label{eq penalty1}\\
  \Delta\Delta G_{x+\Delta x}^{n} &=& -RT \ln\left(\frac{I^{n}_{x+\Delta x}}{I_{PM}}\right) \label{eq penalty2} \\
  \Delta\Delta G_{x,x+\Delta x}^{m,n} &=& -RT \ln\left(\frac{I^{m,n}_{x,x+\Delta
  x}}{I_{PM}}\right) \label{eq penalty3}
\end{eqnarray}
in which the superscript $m$ and $n$ represent the three possible mismatching
nucleotides at location
$x$ and $x+\Delta x$ respectively. If the additivity of the NN model
holds, the free energy penalty of Equation~(\ref{eq penalty3}) should
equal the sum of the individual penalties of Equations~(\ref{eq penalty1})
and (\ref{eq penalty2}). To test this, we introduce
\begin{equation}\label{Eq alpha}
\alpha = \frac{\Delta\Delta G_{x}^{m} + \Delta\Delta G_{x+\Delta x}^{n}
- \Delta\Delta G_{x,x+\Delta x}^{m,n}} {\Delta\Delta G_{x}^{m} +
\Delta\Delta G_{x+\Delta x}^{n}}
\end{equation}
which measures the relative deviation from additivity. Figure~\ref{FIG_interaction_alpha}
shows the experimental results for $\alpha$
in which we averaged over $x$, $m$ and $n$, leaving $\alpha$ as a function
of the distance $|\Delta x|$ between two mismatches.
From this data, we notice that $\alpha$ has a value of about zero
when the mismatches are separated by $\geq 5$ nt, but a
clear positive value for smaller  $\Delta x$. Apparently the free
energy penalty of two nearby mismatches is smaller than the sum of
the two individual contributions, resulting in a positive $\alpha$.
Furthermore, the inset from Figure~\ref{FIG_interaction_alpha} shows
that the relationship is linear in a semi logarithmic plot, hence
$\alpha$ decays exponentially with $|\Delta x|$. Note that at
$\Delta x = 0$ only one mismatch is present, hence $m = n$ and
$\alpha$ will be identical to $1/2$ according to Equation~(\ref{Eq
alpha}). All these observation result from direct measurement
values, containing no fitting parameters and strongly suggest that
in double-stranded DNA, mismatches have a physical interaction with
each other which decays exponentially to zero over a distance of
about five nucleotides.

These results are setting some limitations on the additivity of the
NN model. However, outside this interaction region of $4$ nt
we expect the NN model to hold i.e. $\alpha$ should be
zero and mismatches can be considered as isolated. This can be
explicitly checked in a very direct way. When $\alpha = 0$ we get
from Equation~(\ref{Eq alpha})
\begin{eqnarray}
\Delta\Delta G^{m}_{x} = \Delta\Delta G^{m,n}_{x,x+\Delta x}-
\Delta\Delta G^{n}_{x+\Delta x}.
\label{direct DDG}
\end{eqnarray}
The free energy penalty $\Delta\Delta G^{m}_{x}$ of a mismatch $m$
at location $x$, which we will call the focus mismatch $(m,x)$, can
be estimated either directly using Equation~(\ref{eq penalty1}) or via a
second mismatch $(n,x+\Delta x)$ using Equations~(\ref{eq penalty2}) and
(\ref{eq penalty3}) for any choice of $n$ and $\Delta x > 4$. Hence,
the free energy penalty of the focus mismatch can be estimated from
measurements in many independent ways and they should provide the same
answer if additivity holds. Note that, using Equations~(\ref{eq penalty2})
and (\ref{eq penalty3}) $I_{PM}$ drops out in the right hand side of
Equation~(\ref{direct DDG}).

\begin{figure*} 
\vbox{
\includegraphics[width=1\columnwidth]{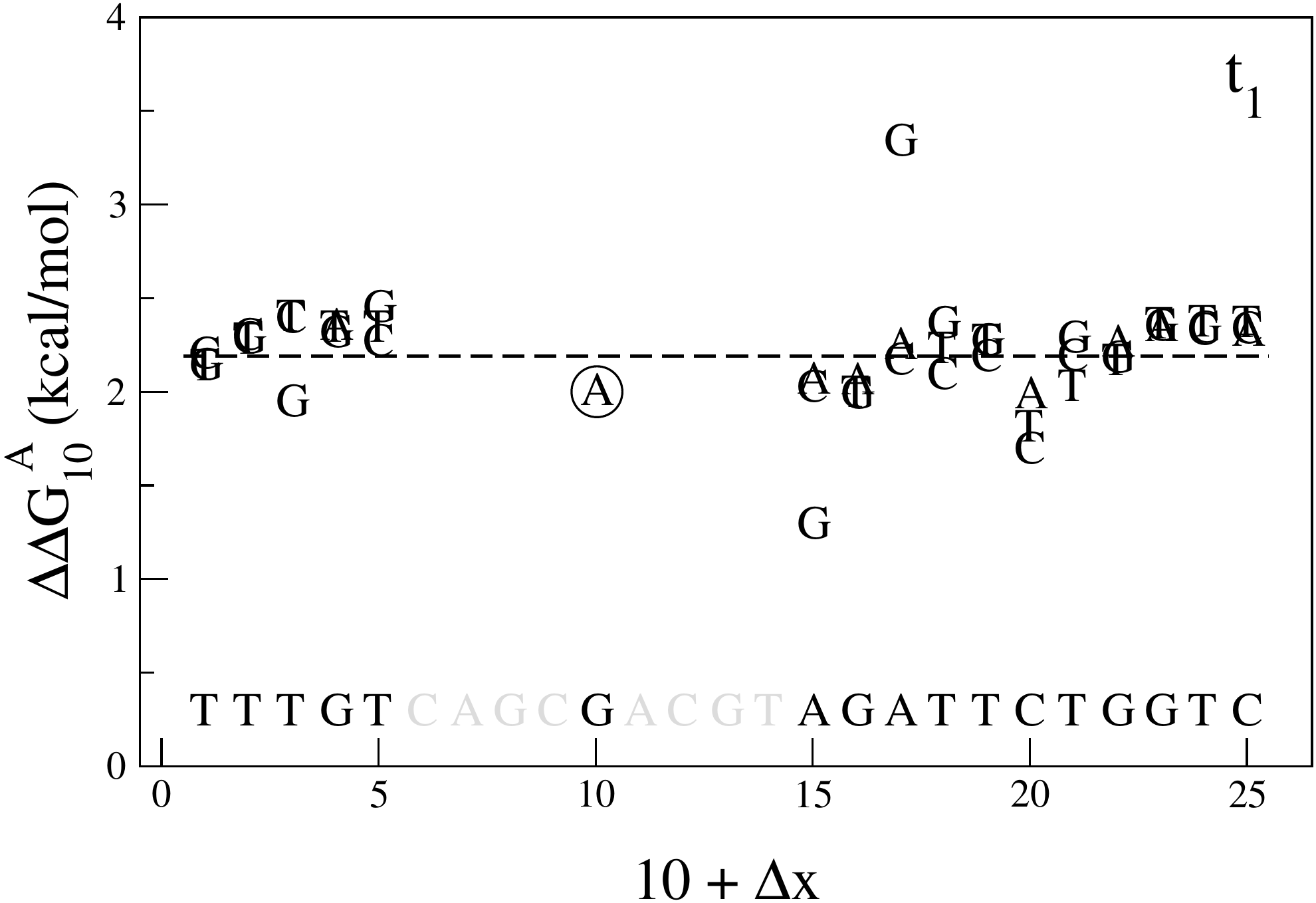}
\includegraphics[width=1\columnwidth]{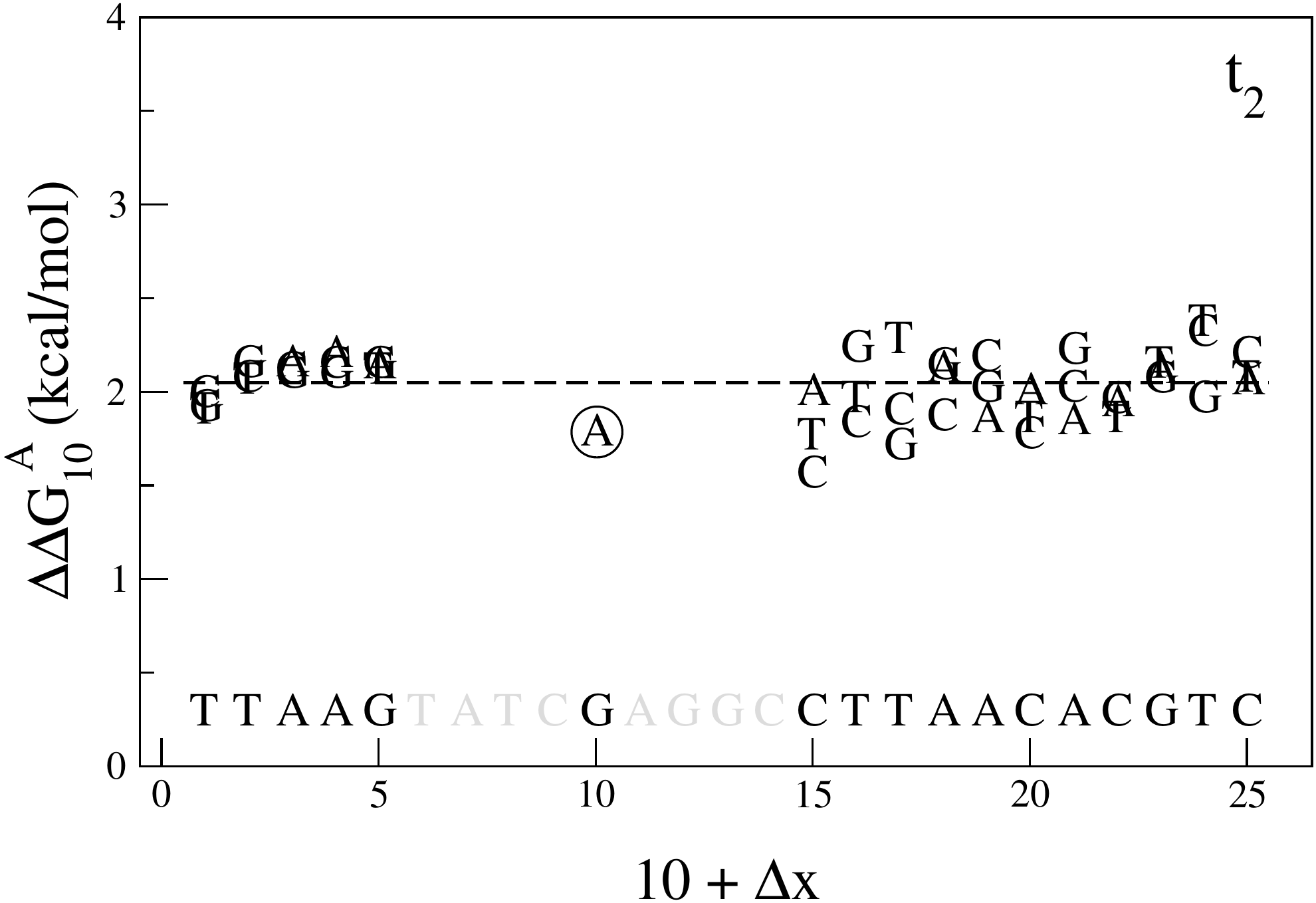}
} \caption{Free energy penalty $\Delta\Delta G^{A}_{10}$ for focus
mismatch $(m=A,x=10)$ derived from experimental intensities
according to Equation~(\ref{direct DDG}) as a function of the location
$x+\Delta x$ of the second mismatch $(n,x+\Delta x)$. For each
$|\Delta x > 4|$ the three values, one per possible mismatch, are
indicated by the letter representing the mismatching nucleotide $n$
of the probe. The target sequence is written in top of the x-axis in
3'-5' notation, $t_1$ in left pane, $t_2$ in right pane. The
dotted line corresponds to the median value of the 48 estimates. The
circled point is the estimate without second mismatch coming from
Equation~(\ref{eq penalty1}). For this particular mismatch, the free energy
penalty for both $t_1$ and $t_2$ is identical and corresponds to
$\Delta \Delta G \left( {G\underline{A}T}\atop{C{\underline
G}A} \right)$.
}
\label{FIG_MEDIAN}
\end{figure*}

Figure~\ref{FIG_MEDIAN} illustrates how Equation~(\ref{direct DDG}) can
be used to estimate the $\Delta \Delta G$ using different combinations of
$n$ and $\Delta x$. In this specific example we consider $\Delta\Delta
G^{A}_{10}$ which corresponds both for target $t_{1}$ and $t_{2}$ to $\Delta
\Delta G \left( \begin{array}{c} {G\underline{A}T}\\{C\underline{G}A}
\end{array}\right)$ (in the Supplementary Data, we show other
examples featuring additivity for different focus mismatches).

In the pane for target $t_2$, all the estimates of the free energy
penalty are close the each other, the $48 + 1$ estimates tightly lie
around a median value, in this case $\sim 2.1$ kcal/mol, indicated
by the dotted line. The picture in the right pane is a typical one
which we observe for any focus mismatch $(m,x)$. This confirms that
additivity holds in the regime $\Delta x>4$, i.e. when mismatches
are separated by $> 4$ nt. Moreover, it shows that
the microarray measurement is internally consistent. Secondly, the
left pane, i.e. experiment $t_1$, provides the same median value for
the free energy penalty, showing also the robustness of the
microarray approach to estimate free energies of DNA structures.
However, this figure was chosen because it is atypical in the sense
that one notices two pronounced outlying values. They correspond to
a sequence where both the focus mismatch and the second mismatch are
of type $AG$. Since they clearly deviate from an otherwise nicely
consistent picture, we believe there must a physically underlying
reason for it. We will come back to this point in the section
where we discuss thermodynamic outliers.

Note that with this second method we accessed values for the free
energy penalties of isolated mismatches without using any multiple
regression or fitting procedure, but we simply compared the
ratios of intensities, Equations~(\ref{eq penalty1})-(\ref{eq penalty3}),
to get a consistent set of independent estimates. The free energy
penalties are then obtained from the median over all data points.
We compared the free energy penalties obtained from this method (median)
with those obtained from linear regression as discussed in the previous
section. The two sets of data are well-correlated
with a Pearson correlation coefficient equal to $0.966$
(see Supplementary Data). This correlation
shows the equivalence of the two approaches. In this analysis, we
restricted ourselves to mismatches in the bulk of the sequence, i.e.
 $x$ is $>5$ nt from the border. Closer to the
border we observe boundary effects, which are covered in the next section.

\subsection{Boundary effects}

The previous section ended by showing the equivalence of both
approaches to access free energy penalties of an isolated mismatch,
provided the data are restricted to bulk mismatches. The direct
median method of the previous section can also assess penalties of
mismatches close to the boundary, whereas on the contrary the fitting
method cannot by construction. The latter, however, has the advantage
of fitting a full parameter set of the NN model and as such can
easily provide bulk values for the free energy penalty of any
isolated mismatch. The combination of both methods now provides an
elegant way to assess the effect of boundary proximity on an
isolated mismatch. Hereto, we introduce the parameter $\beta$ as the
relative reduction of free energy penalty of a mismatch when
compared to its bulk value.
\begin{eqnarray}
\beta = \frac{\Delta\Delta G_x^m}{\Delta\Delta G_{bulk}^m}
\label{equation beta}
\end{eqnarray}
\begin{figure}
  \includegraphics[width=1\columnwidth]{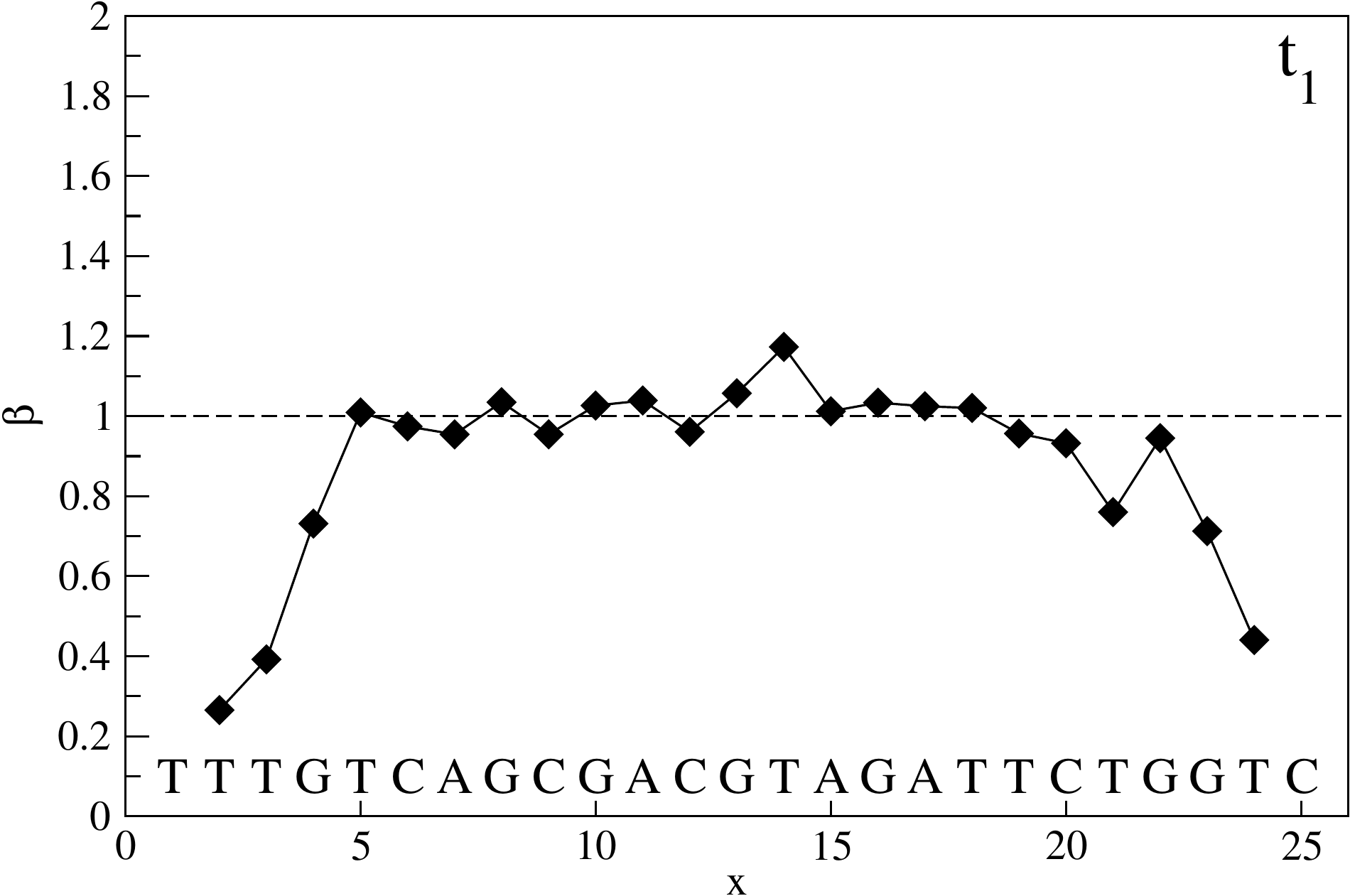}
  \caption{Boundary effect: $\beta$, the relative reduction of mismatch free
energy penalty, as a function of location for experiment with target
$t_1$. Each point is the average of three estimates, one per
possible mismatch. Data are absent for the extremal locations $x=1$
and $x=25$, since no value can be calculated by the NN
model.}\label{FIG Boundary}
\end{figure}
In Figure~\ref{FIG Boundary} the parameter $\beta$ is shown as a
function of $x$ after averaging over $m$. It is clear that, as
expected, $\beta$ is approximately equal to one in the bulk, whereas when
approaching the boundary, a reduction of free energy penalty occurs
which reaches up to 80\%. Note that for mismatches at the
boundary, $x=1$ and $x=25$, the NN model is not applicable and no
data is presented. Figure~\ref{FIG Boundary} show that the range
of the boundary effect is $\sim 4$ nt.

\subsection{Thermodynamic outliers}
\label{sec:thermodynamic outliers}

As a final result of this article, we come back to the two outliers
observed in Figure~\ref{FIG_MEDIAN}(a); the same deviations are found
in replicated experiments at different concentrations: therefore,
they are unlikely due to experimental errors. For these two cases we
find $\Delta\Delta G^{A,G}_{10,15}-\Delta\Delta G^{G}_{15}\approx
1.2$ kcal/mol and $\Delta\Delta G^{A,G}_{10,17}-\Delta\Delta
G^{G}_{17}\approx 3.1$ kcal/mol, strongly deviating from the median
value ($\approx 2.1$~kcal/mol).  The common feature of these two
sequences is that they involve GA mismatches. The two set of
mismatches are arranged in an antiparallel way i.e. one G and one A
are on the same strand. Mismatches of GA type in DNA and RNA helices
have been the subject of several studies in the past~
\cite{Li1991,Li1991a,SantaLucia1993,LANE1994,EBEL1994,Morse1995,Seela2008,Chang2011a}. In the
RNA folding, it is known that GA pairs contribute substantially to
the RNA helix stability. Their contribution is comparable to that of
a canonical AT pair. As AT pairs, GA form two hydrogen bonds, but
can also assume four different conformations~\cite{Li1991}.
The microarray data suggest that the antiparallel
combination of GA and AG pairs of mismatches have a long range
interaction effect, which is probably a signature of some structural
conformational change of a double helix containing these pairs.
Next-nearest neighbor effects extending up to $4$ nt
distance for antiparallel GA mismatches have been reported in the
case of RNA duplexes in~\cite{Morse1995} (longer distances were
not considered that case).  We investigated
antiparallel GA and AG pairs of mismatches also in sequences $t_2$
and $t_3$, but found no anomalous behavior in those cases.
This suggests that the nucleotide sequences between the two $GA/AG$
pairs plays an important role in the overall stability of the duplex.
\begin{figure}
\vbox{
\includegraphics[width=1\columnwidth]{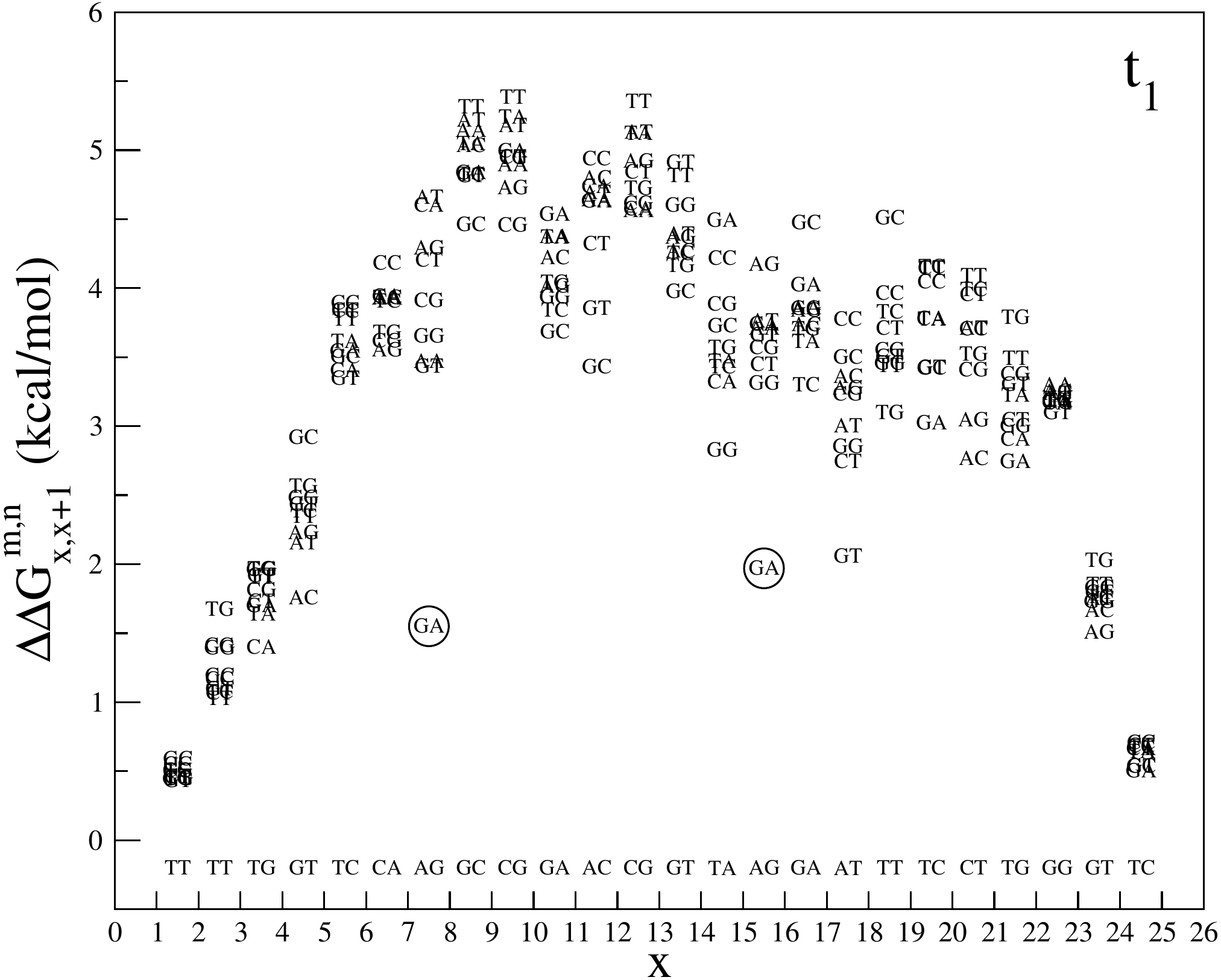}
} \caption{The free energy penalty of tandem mismatches, from
experiment with target $t_1$:
$\Delta\Delta G \left(
      x\underline{mn}y \atop
      x'\underline{ab}y'
\right)$, where $x'$ and $y'$ are complementary to $x$ and $y$
respectively. $ab$ denoted above the x-axis are
the fixed nucleotides in the target. $mn$ is
a tandem mismatch in the probe and the vertical position of these
letters in the plot give the associated free energy penalty.
Note the low free energy penalty for
$
      5'-\underline{GA}-3' \atop
      3'-\underline{AG}-5'
$
mismatches (encircled).
} \label{FIG_tandem}
\end{figure}

As a further proof of the outlying behavior of antiparallel $GA/AG$
pairs we show in Figure~\ref{FIG_tandem} a plot of free energy
penalties for tandem mismatches (neighboring double mismatches).
These are again obtained from Equation~(\ref{eq penalty3})
for different $m$ and $n$ mismatching nucleotides, where in
the case of tandem mismatches, $\Delta x$ is equal to 1. On each
location of the sequence our data set contains nine different types of
tandem mismatch. A clear boundary effect is noticeable, but when
looking at the bulk data points tandem mismatch of the type $GA/AG$
are again outlying, they appear to be particularly stable with a
free energy penalty $\sim 2$ kcal/mol below average.

\section{DISCUSSION AND CONCLUSION}

In this article, we have analyzed DNA hybridization reactions in
microarrays and quantified free energy penalties of single and double
mismatches. We have shown that the experimental data are very
precise and reproducible. The microarray data follow an equilibrium
isotherm over a range of four orders of magnitude in the
fluorescence intensities and allow the extraction of accurate
thermodynamic parameters. First, the analysis provides a database
with a large number of NN parameters for isolated mismatches. These
parameters correlate well with those reported in the literature from
hybridization experiments in solution. Second, the experiments
contain systematic measurements of hybridization with two
mismatches, which allowed us to probe the validity limit of the NN
approximation. We showed that when two mismatches are separated by a
distance of $\geq 5$ nt their effect is additive,
allowing a standard approach with the NN model. However, for shorter
distances, the additivity is no longer valid and we found that
duplexes with neighboring mismatches are more stable than expected
from additivity. This interaction was shown to decay exponentially
as a function of the distance between mismatches. Further, we
investigated the behavior of mismatches close to the helix edges,
and showed that their free energy penalty is reduced up to $80\%$
when compared to the bulk behavior. The boundary effect
was observable up to $4$ nt from the helix edge.
Finally, we also found some thermodynamic outliers, sequences
involving two antiparallel GA mismatches, in which the mismatch
interaction appears to persist beyond $5$ nt. These
outliers were not related to experimental error indicating a
signature of some structural conformational change of a double helix
containing these mismatch pairs.

Overall, the analysis of the microarray data reported in this article
provides new quantitative insights on the DNA hybridization
parameters, on the NN model and its present limitations. Our study
is in line with a number of recent articles, which have been dedicated
to the investigations of fundamental physico-chemical properties of
DNA arrays~\cite{Trapp2011,Golovkin2009,Pozhitkov2010,Irving2010,Fuchs2010,
Burden2010a,Qamhieh2009,Gharaibeh2010,Lee2007,Binder2005}.
Due to the relevance of hybridization in many technologies,
going from PCR~\cite{albe02} to recent developments in
biosensors, e.g.~\cite{Grinsven2011}, a good thermodynamic model is also
important from the application point of view. A precise
quantification of interaction free energies involved in the
hybridization will help to increase the accuracy of microarrays and other
hybridization-based technologies, so that these devices could realize
their full potential, for instance, for clinical applications~\cite{Hooyberghs2010a}.
For these applications, an increase in specificity and sensitivity is very
important and can be achieved through better understanding of fundamental
properties of hybridization in these devices.

There has been considerable attention in recent years~\cite{Hooyberghs2009,Trapp2011,Pozhitkov2010,Irving2010,Fuchs2010,Burden2010a,
Qamhieh2009,Gharaibeh2010,Lee2007,Binder2005,Grinsven2011,Hooyberghs2010a,Shanahan2012}
in understanding the fundamentals of hybridization in DNA microarrays
and its impact in data analysis. Here, we have shown that microarrays 
are a reliable and high-throughput tool to gain insight on DNA hybridization
thermodynamics. The same method could be used to screen other types of
defects, as bulges. Indeed, it was recently used for understanding loop conformations~\cite{Trapp2011}.

\section{SUPPLEMENTARY DATA}
Supplementary Data available in Appendix.

\section{ACKNOWLEDGEMENTS}
We thank Karen Hollanders for expert technical assistance. We acknowledge
financial support from Research Foundation-Flanders (FWO) Grant No. G.0311.08,
KULeuven Grant No. STRT1/09/042 and VITO Grant ZL39010200-401.\newline

\noindent Conflict of interest statement: \textit{None declared}.

\newpage
\appendix

\textbf{SUPPLEMENTARY DATA}
\section{Nearest neighbor model and linear regression}
\label{subsec:nn_model}

According to the nearest-neighbor model, the total hybridization
free energy of a target to a probe can be expressed as a sum of the
dinucleotide parameters $\Delta G_{\alpha}$ accounting for hydrogen
bonding and stacking interactions. The index $\alpha$ covers all
possible dinucleotide parameters. Some examples are:
\begin{eqnarray}
\Delta G &&\left( {5'-AT-3'}\atop{3'-TA-5'} \right), \nonumber\\
\Delta G &&\left( {5'-AC-3'}\atop{3'-TG-5'} \right), \nonumber\\
\Delta G &&\left( {5'-A{\underline A}-3'}\atop{3'-T\underline{A}-5'} \right)
\label{dinucl}
\end{eqnarray}
where the underlined nucleotides indicate mismatches. In total there
are $10$ perfect match parameters (taking into account symmetries)
and $48$ parameters in the case of a single mismatch. These
dinucleotide parameters are known not to be unique, see e.g.
\cite{gray97a}.

Thermodynamics predicts that the intensity measured from a spot $I_i$
is given by:
\begin{equation}
I_i = I_0 + A c e^{-\Delta G_i/RT}
\label{IivsDG_supp}
\end{equation}
where $\Delta G_i$ is the total hybridization free energy between a
target and a probe, $A$ is a parameter which sets the intensity
scale, $c$ the target concentration, $R$ the gas constant and $T$
the temperature. $I_0$ is the aspecific signal that can be
considered as background. In this paper the stability of duplexes
was always compared to that of the perfect match, i.e.
\begin{equation}
y_i \equiv \ln I_i - \ln I_{PM} =
-\frac{\Delta G_i - \Delta G_{PM}}{RT}
\end{equation}
which defines the free energy penalty of probe $i$ with respect to
the perfectly matching probe. This penalty can be expressed as a sum
of nearest-neighbor dinucleotide parameters:
\begin{equation}
y_i = \sum _ {\alpha = 1} ^ { 58 } X _ { i \alpha }
\frac{\Delta G_{\alpha}}{RT}
\label{lin_prob_suppl}
\end{equation}
where $X_{ i \alpha }$ is the frequency matrix, which counts the
number of times a given dinucleotide term contributes to $y_i$. As
an example, for an isolated mismatch of type GA we have:
\begin{eqnarray}
\Delta G_i \left( {\ldots G\underline{A}T \ldots}\atop{\ldots
C{\underline G}A \ldots} \right) &-& \Delta G_{PM} \left( {\ldots
GCT \ldots}\atop{\ldots CGA \ldots} \right) =
\nonumber\\
\Delta G \left( {G\underline{A}}\atop{C{\underline G}} \right) &+&
\Delta G \left( A{\underline{G}}\atop{T{\underline A}} \right)
\nonumber\\
- \Delta G \left( {GC}\atop{CG} \right) &-& \Delta G \left(
{CT}\atop{GA} \right)
\nonumber\\
\equiv \Delta \Delta G \left( {G\underline{A}T}\atop{C{\underline
G}A} \right) && \label{yi_eq2_supp}
\end{eqnarray}
For notational convenience we used, by symmetry, the equality of
$\Delta G \left( {\underline{A}T}\atop{{\underline G}A} \right)
 = \Delta G \left( A{\underline{G}}\atop{T{\underline A}} \right) $ to
have the mismatch on the right hand side of the dinucleotide. For
any given $i$, the matrix elements $X_{i\alpha}$ are all zero except
for the four dinucleotide terms of Equation~(\ref{yi_eq2_supp}) which
contribute by $+1$ for the two dinucleotides with mismatches and
$-1$ for the two perfect matching dinucleotides.
Equation~(\ref{lin_prob_suppl}) defines a multiple linear
regression, from which the $58$ dinucleotide parameters can be
fitted to match all the observed free energy penalties of
mismatches. Note that it defines the dinucleotide parameters not in
a unique way, e.g. the following transformation
\begin{eqnarray}
\Delta G
\left( {x\underline{A}}\atop{x'{\underline G}} \right)
&\to&
\Delta G \left( {x\underline{A}}\atop{x'{\underline G}} \right)
+ \varepsilon
\label{shift1_suppl}\\
\Delta G
\left( {x\underline{G}}\atop{x'{\underline A}} \right)
&\to&
\Delta G \left( {x\underline{G}}\atop{x'{\underline A}} \right)
- \varepsilon
\label{shift2_suppl}
\end{eqnarray}
in which the same constant $\varepsilon$ is added and subtracted to
different dinucleotide parameters, leaves Equation~(\ref{yi_eq2_supp})
invariant. The triplet parameters, such as defined in the last line
of Equation~(\ref{yi_eq2_supp}) are however unique as expected, since
they are directly physically accessible.

\section{Target sequence selection with Optimal design}
\label{subsec:odesign}

As discussed above, the dinucleotide parameters can be obtained from
a linear fit from $N$ independent experimental measurements. Such an
approach always contains some uncertainties. These uncertainties can
be lowered if one takes $N$ large. In our specific case $N$ equals
the number of spots on the microarrays, and can be increased by
combining data from more arrays (see main paper for experimental
setup).  Further, for a given fixed value of $N$ one can use some
optimization criterion to select the best $N$ measurements which
minimize the uncertainties on fitted parameters. In our case this
comes down to the selection of a target sequence with good
statistical properties. The theory of {\sl Optimal Design}
establishes some criteria for this purpose and we briefly discuss
this theory here.

Before entering into the details of the optimization followed in the
microarray experiment we discuss a one dimensional example, which
illustrates the optimization method. Let us take the example of a
simple linear regression with an intersect set to zero
(corresponding to a one-dimensional system):
\begin{equation}
 y_i=\beta x_i\,,
\end{equation}
where $\beta$ is the unknown of the problem, $x_i$ and $y_i$ are
respectively the input and output of the experiment $i$ and can take
any real value. The parameter $\beta$ can be obtained by the least
square method~:
\begin{equation}
\beta=\frac{\sum(x_i-\bar x)(y_i-\bar y)}{\sum(x_i-\bar x)^2}\,,
\end{equation}
where the symbol $\bar{\dots}$ means the average over the $N$ elements.
The error on $\beta$ is given by~:
\begin{equation}
\Delta \beta\underset{N\gg1}{=}
\frac{S}{N}\sqrt{\frac{1}{\sum(x_i-\bar x)^2}}\,,
\label{costfunc}
\end{equation}
where $S$ is the cost function of the system. Equation~(\ref{costfunc})
implies that the error can be decreased by enlarging the sampled
points ($N$) or, for $N$ fixed, by increasing the variance of the
variable $x_i$. The latter criterion can be used in the design of
the experiment by performing measurements $y_i$ for a well spread
set of points $x_i$. Indeed, it is intuitively clear that when $x_i$
are very close to each other (small variance) one has a large
uncertainty on the estimate of the slope $\beta$. In what follows we
discuss about optimal design criteria in higher dimensions, which
roughly correspond to the idea of the maximization of the variance
in the previous one-dimensional example.

We define first the so-called {\it information matrix} $M = X^TX$,
where $X$ is the frequency matrix defined in Equation~(\ref{lin_prob_suppl})
and where $X^T$ denotes its transpose. In terms of matrix elements:
\begin{equation}
 M_{\alpha \beta} = \sum_{i=1}^N X_{i \alpha} X_{i \beta}
\end{equation}
which is thus in our case a square symmetric matrix of dimension $58
\times 58$.

The information about the quality of the experimental design is
encoded in $M$ and in our case is defined by the sequence of the
target oligo in the experiment (see main paper for experimental
setup). The three most used criteria in optimal design are the A-,
D- and E-optimality. A-optimality corresponds to minimizing the
trace of $M^{-1}$, D-optimality corresponds to minimizing the
determinant of $M^{-1}$ and E-optimality corresponds to maximizing
the lowest eigenvalue of $M$. Roughly speaking, these strategies
amounts to maximize the information encoded in $M$ ~\cite{atki92}.
We note that in the linear problem of Equation~(\ref{lin_prob_suppl}) the
information matrix has a minimum of $7$ null eigenvalues (see the
supplementary material of Ref.~\cite{hooy09} for a detailed
explanation). These come from unavoidable degeneracies of the
problem, or equivalently from the fact that the dinucleotide
parameters are not unique (see e.g. Equations~(\ref{shift1_suppl}) and
(\ref{shift2_suppl})). Having some zero eigenvalues, the information
matrix $M$ is not invertible, therefore we are working with
pseudo-inverse which is obtained from the singular value
decomposition of $M$.

The three target sequences, $t_1, t_2$ and $t_3$, which were used
for the experiments and which are mentioned in table 1 of the main
article were selected as follows. We collected a set of candidate
targets by scanning over a piece of the human genome and taking
subsequences of length 25. The first criterion was to choose
sequences with minimum, unavoidable, number of 7 zero eigenvalues
in order to get the minimum number of degeneracies when solving the
linear system to estimate the nearest-neighbor parameters, as discussed
above. For $t_1$, we considered a subset of sequences with a minimum
distance of 3 nucleotides from the border and a minimum distance of 3 
nucleotides between 2 mismatches. For $t_2$ and $t_3$, the minimal distance from
the border is 4 nucleotides and the distance between 2 mismatches
is at least 5 nucleotides. Therefore, the constraint on the subset to select
$t_1$ was weaker than $t_2$ and $t_3$. Since the constraint for $t_2$ and $t_3$ is
stronger than $t_1$, in this case, the number of equations in the linear system is
lower and it is more difficult to find subsequences of length 25 which
display the minimum number of 7 zero eigenvalues. For the same order of
calculation, we managed to find 130 sequences for $t_1$ and only a
few sequences for $t_2$ and $t_3$.  For $t_1$, this set of candidates was
subsequently ranked according to the three optimal design criteria A, D and E.
Finally, the candidate targets which ended up as top-ranked on all three
criteria were retained. Moreover, we checked the energy for the target
to fold on itself. For the 3 targets, it takes a reasonable value.


\section{The linear regime}
\label{subsec:linear_regime}

\begin{figure}
\includegraphics[width=0.8\columnwidth]{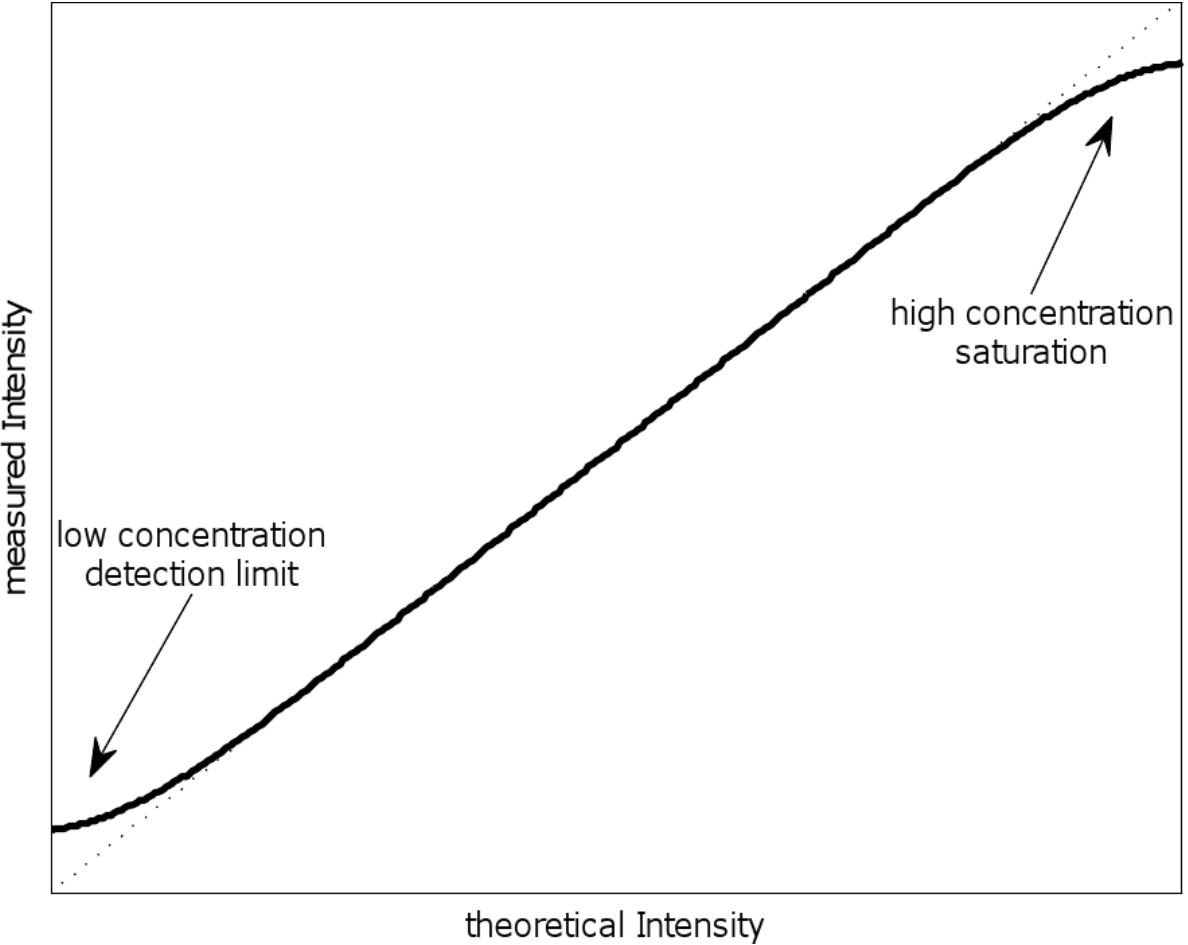}
\caption{Sketch to show non-linear behaviour due to detection limit
on low end and saturation on high end.} \label{FIG_linear_sketch}
\end{figure}
\begin{figure}[h]
\includegraphics[width=1\columnwidth]{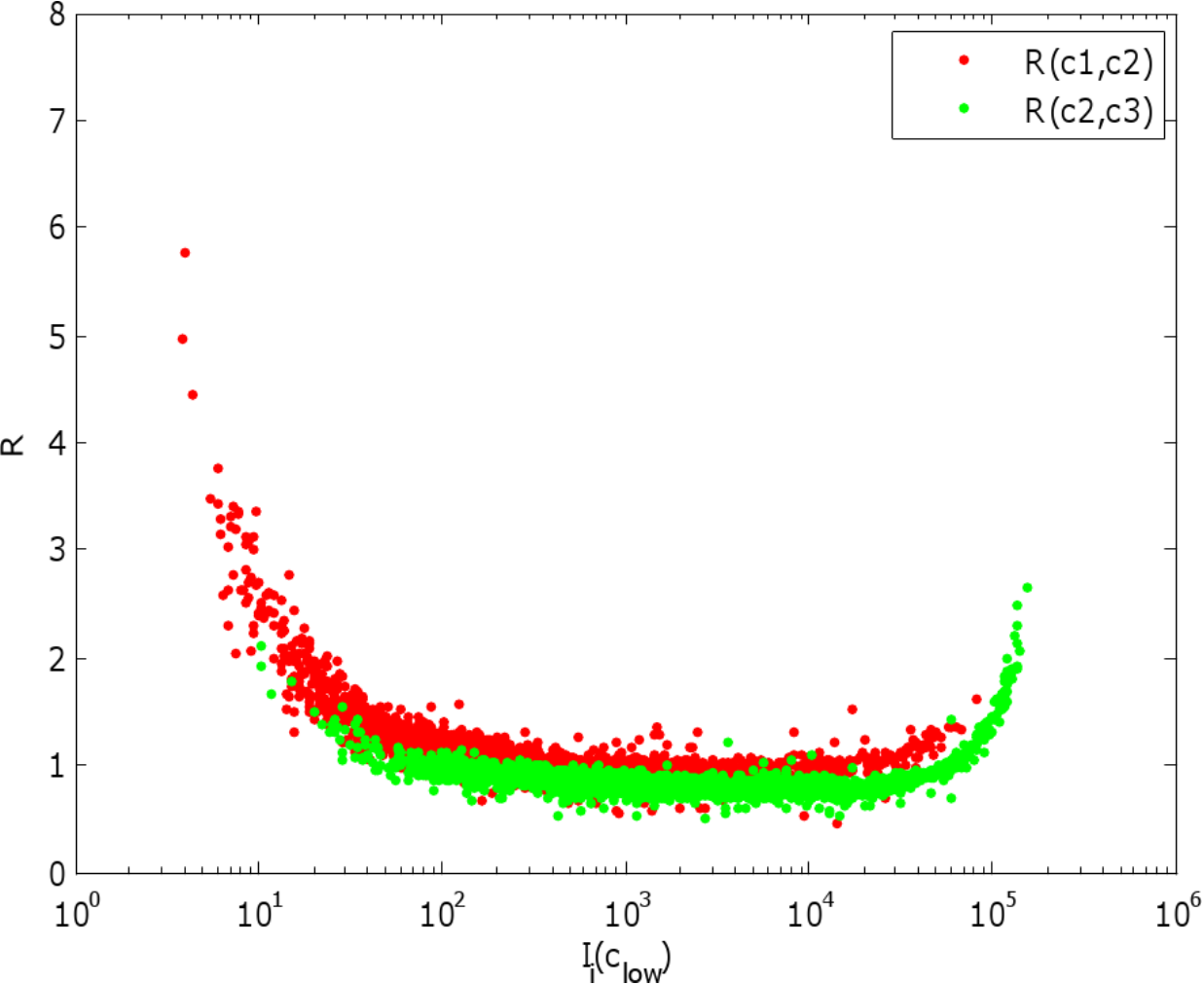}
\caption{Plot of $R_i(c_n,c_{n+1})$, $c_{n+1}>c_n$, as a function of
the Intensity $I_i(c_n)$. The used target concentrations are $c_1=20
pM$, $c_2=100 pM$, $c_3=500 pM$ of target $t_2$. } \label{FIG_linear_regime}
\end{figure}

As a measurement device the microarray technology is faced with a
detection limit in the low measurement regime and a saturation in
the high end: see sketch in Figure~\ref{FIG_linear_sketch}. In our
research we want to limit ourselves to measurements in the linear
regime. To assess which data meet this requirement we combine
experiments which are identical (identical target sequence,
identical probe sets, identical hybridisation conditions) except for
the concentration $c_n$ of the target. If the data is in the linear
regime we expect the intensity of a spot $i$ of the experiment with
target concentration $c_n$ to be
\begin{equation}
{I_i \propto c_n exp(-\Delta G_i/RT).} \label{I_i}
\end{equation}
If we now combine two experiments, one with target concentration
$c_n$ and one with $c_{n+1}>c_n$, and define for each spot $i$ the
quantity $R$ as
\begin{equation}
{R_i(c_n,c_{n+1})=\frac{I_i(c_n)}{I_i(c_{n+1})} \frac{c_{n+1}}{c_n}}
\label{R}
\end{equation}
than we expect $R$ to be equal to $1$ when both intensities are in
the linear regime. However for low $c_n$ the spot intensity
$I_i(c_n)$ can be close to detection limit and consequently be
higher than predicted by the theory of Equation~(\ref{I_i}), or for
high $c_{n+1}$ the intensity $I_i(c_{n+1})$ can be close to
saturation and consequently lower than theoretically expected. In
both cases $R$ will be above one. The result of this analysis
is shown in Figure~\ref{FIG_linear_regime} for the combinations
$(c_1=20pM,c_2=100pM)$ and $(c_2=100pM,c_3=500pM)$ of target $t_2$. From this
picture it is clear that for a large part of the intensity range $R$
equals one and supports the linear regime. For the green dots, there
is a deviation in the high intensity range due to the proximity of
saturation of these spots in the $500 pM$ experiment. For the red
dots a deviation is present due to proximity of the detection limit
for these spots in the $20 pM$ data. This approach gives a criterion
to assess the validity of the linear regime per spot and the
possibility to make a correction for the non-linear behaviour close
to saturation or detection limit.


\section{Free energy additivity of mismatches}
\label{subsec:additivity}

\begin{figure*}[h!]
\vbox{
\includegraphics[width=1\columnwidth]{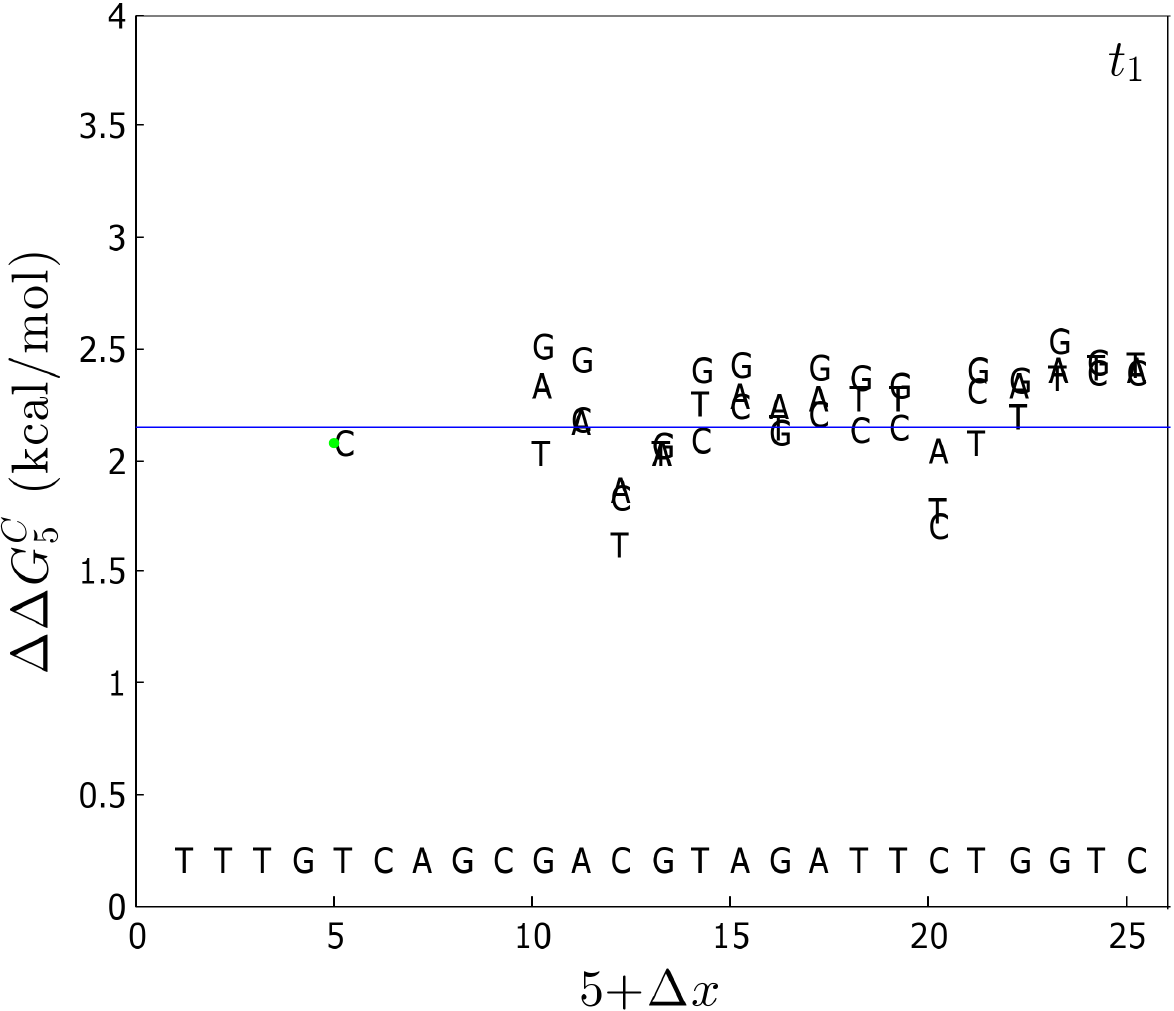}
\hspace{3mm}
\includegraphics[width=1\columnwidth]{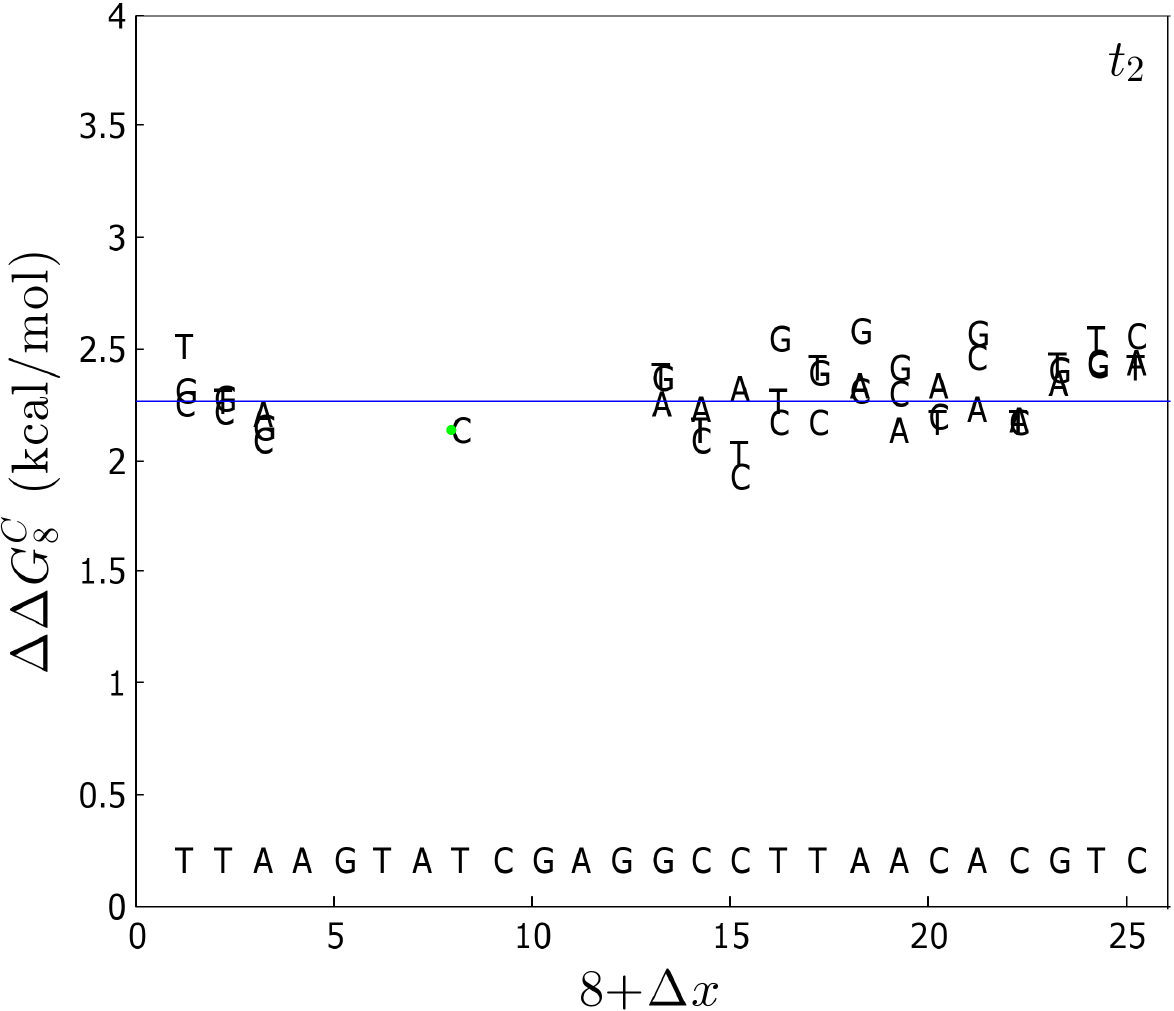}
\vspace{2mm}
\includegraphics[width=1\columnwidth]{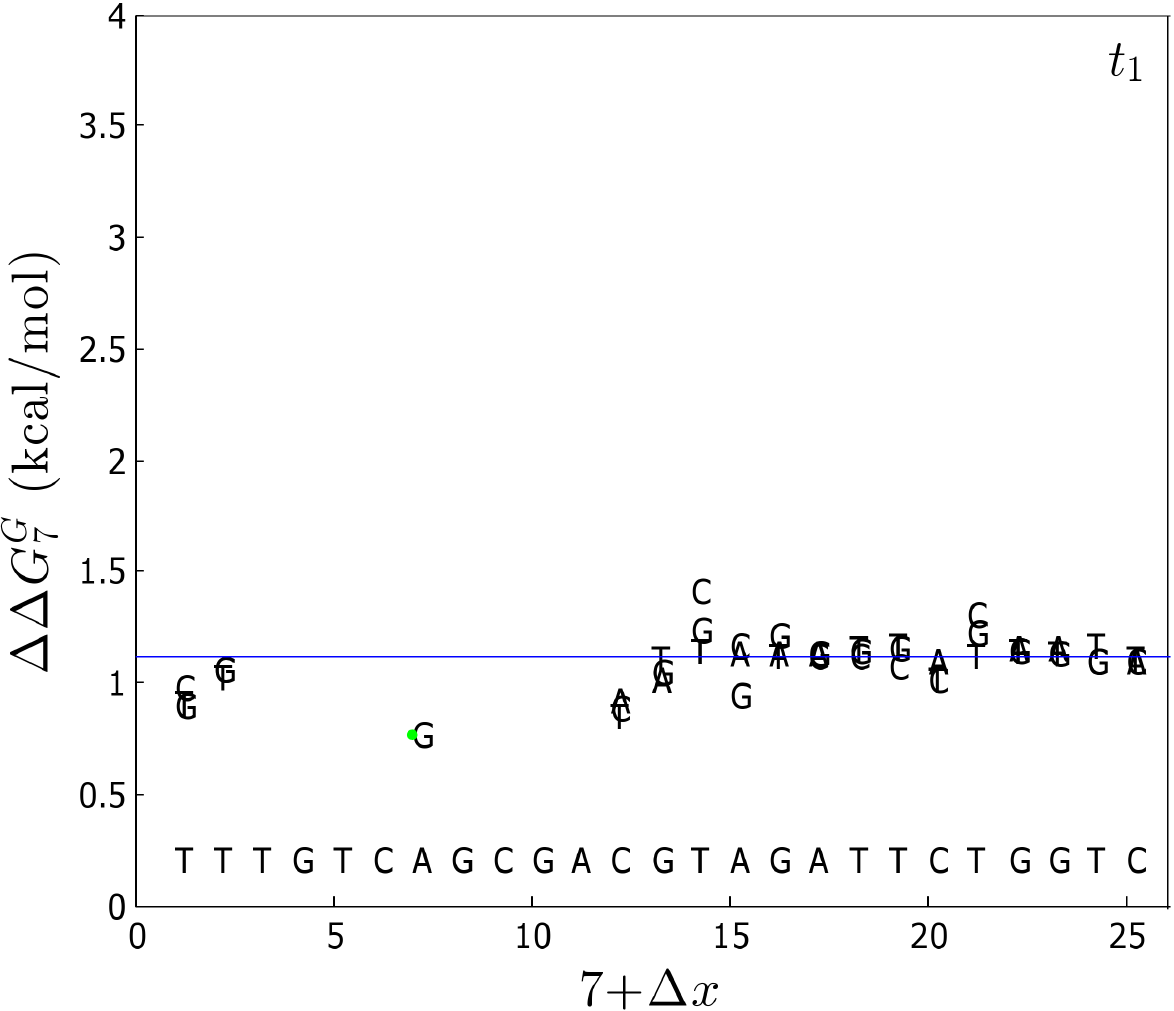}
\hspace{3mm}
\includegraphics[width=1\columnwidth]{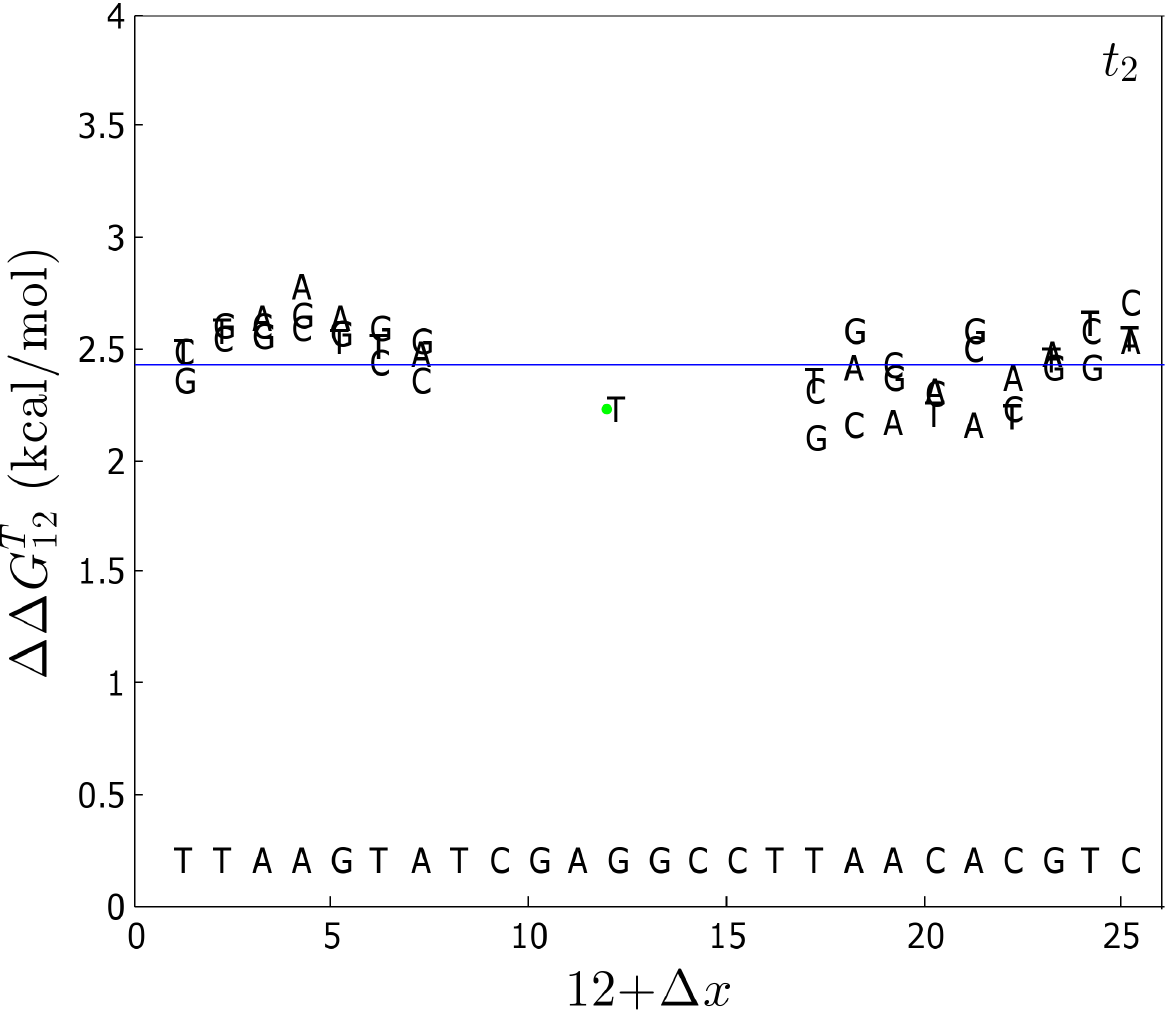}
\vspace{2mm}
\includegraphics[width=1\columnwidth]{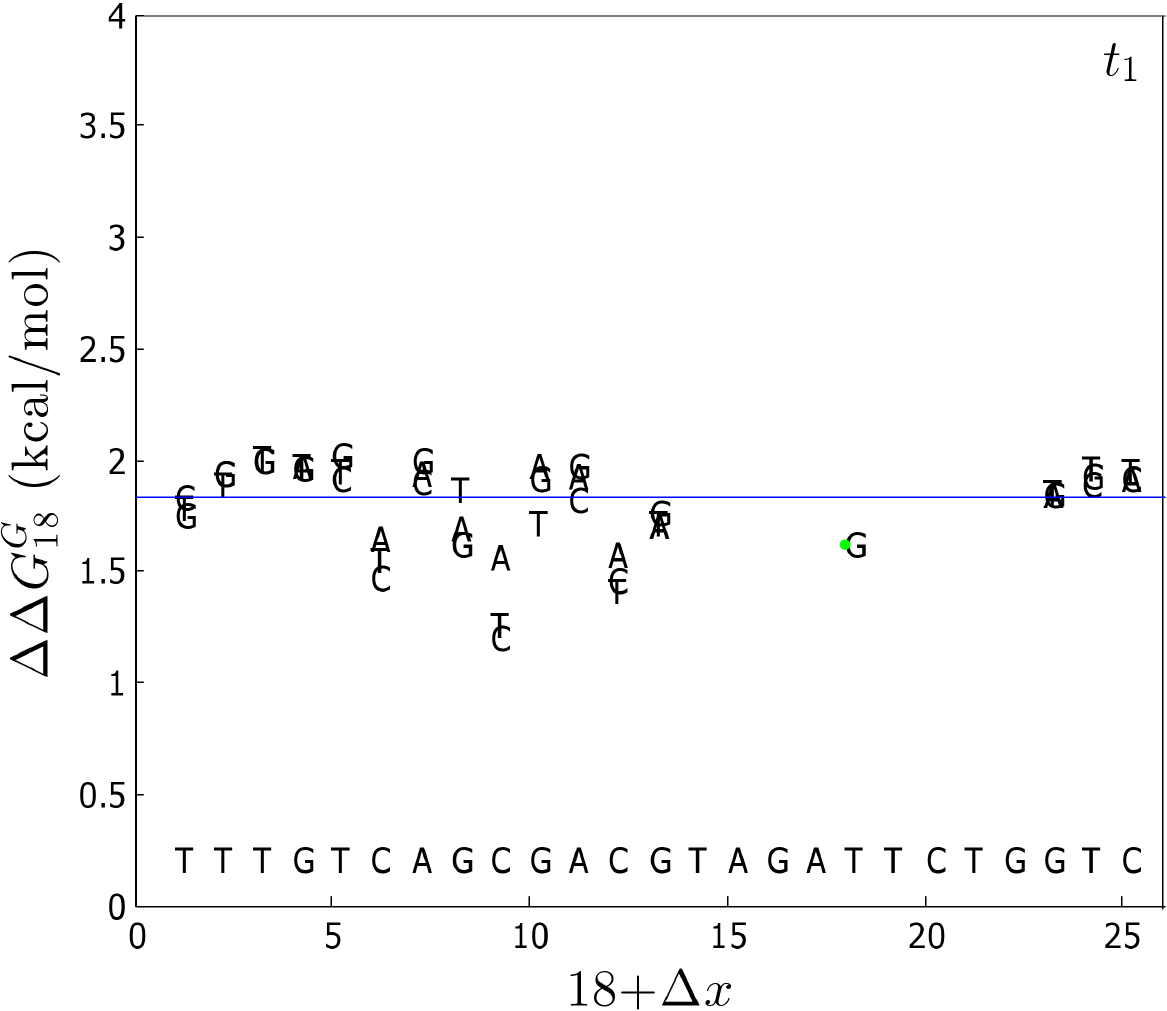}
\hspace{3mm}
\includegraphics[width=1\columnwidth]{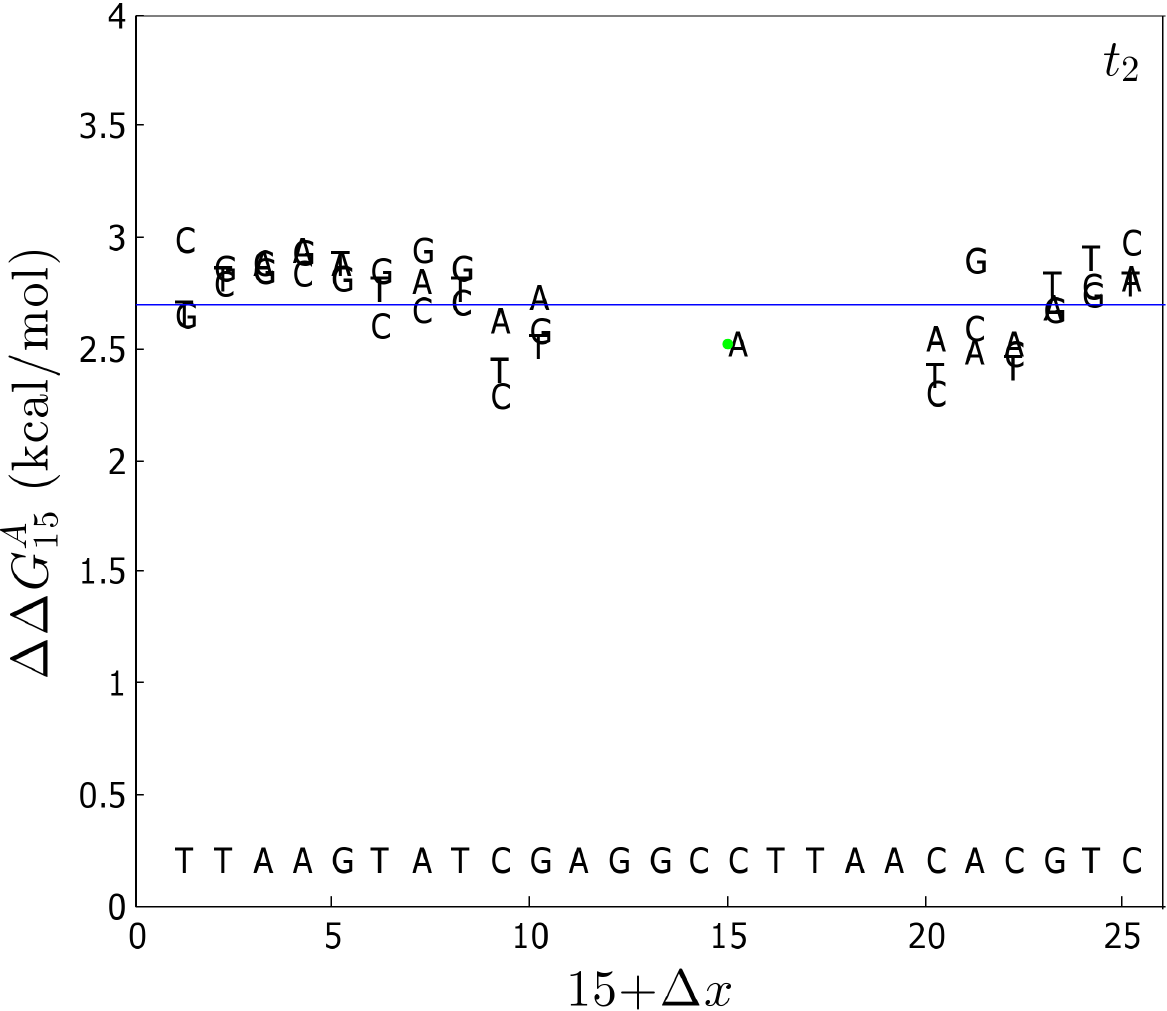}
}
\caption{A few examples of different focus mismatches showing additivity
as $|\Delta x > 4|$. Similar to Figure 5 in the main paper, the target
shown in top of the x-axis is in 3' to 5' orientation,
$t_1$ are on the left side, $t_2$ are on the right side.} \label{FIG_additivity}
\end{figure*}

In the main article the additivity of free energy penalties of
mismatches was shown when mismatches were separated by more than
four nucleotides. For two examples, this was explicitly shown in
Figure 5 of the main article. In this section we add some further
examples of the additivity with similar plots. These are
shown in Figure~\ref{FIG_additivity}.

\begin{figure}[h!]
\includegraphics[width=1\columnwidth]{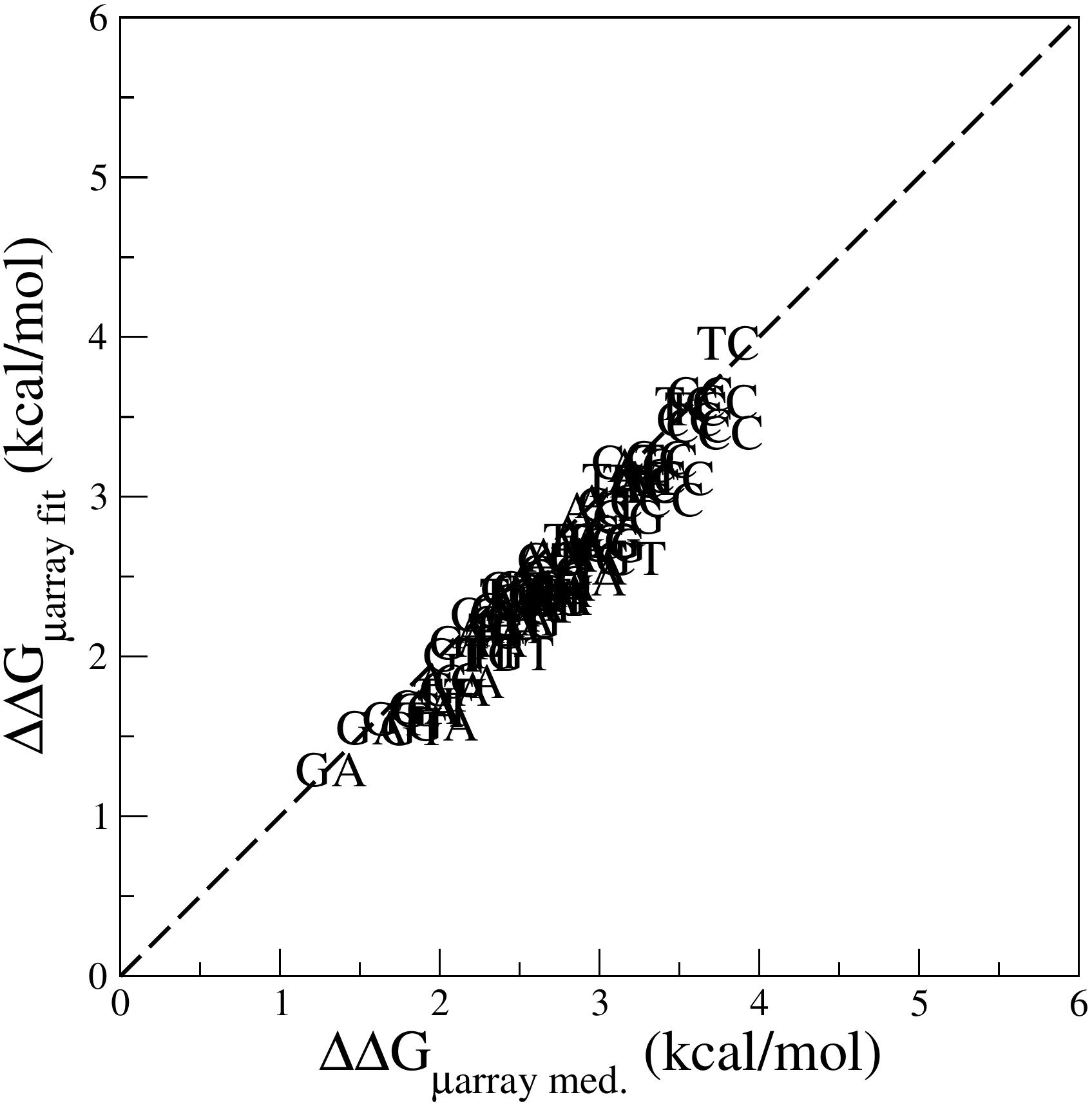}
\caption{Comparison of estimates of free energy penalties for
isolated mismatches, obtained in two different ways: from the linear
model fit and from the median of independent estimates.
The two sets of data are strongly correlated (Pearson's correlation
$0.966$). }
\label{FIG_MEDIAN_VS_FIT}
\end{figure}

\section{Self-consistency in free energy penalties estimation of triplet nucleotides}
In the main article we present two different approaches that can be used
to estimate free energy penalties of single mismatches in a triplet of
nucleotides such as in Equation~(3) of the main article.
The first method, i.e. by linear fitting, produces a robust
estimation provided that each of the 58 NN dinucleotide parameters are equally
well-represented. This was achieved by the use of Optimal Design principle in designing 
the experiments. Another method is by taking the median of data points from
ratios of intensities following Equations~(4)-(6) of the main article.
Figure~\ref{FIG_additivity} of this document shows six of these unique triplets
in which the free energy penalties are indicated by the horizontal line from taking
the median of each independent estimates. It is then imperative to see if 
these two methods are equivalent in providing the estimates.
Figure~\ref{FIG_MEDIAN_VS_FIT} shows that the free energy penalties calculated
from the two methods are well-correlated with Pearson correlation $0.966$
(such as mentioned in the main article). This is indicating the equivalence of
the two methods. This is also a proof that our experiments are self-consistent
from the different perspective of these two approaches.

\end{document}